\documentclass[acmlarge,authorversion,nonacm]{acmart}

\usepackage{adjustbox}
\usepackage{multirow}
\usepackage{xcolor,colortbl}
\usepackage{amsmath}
\usepackage[most]{tcolorbox}

\graphicspath{{figures/}}

\AtBeginDocument{%
  \providecommand\BibTeX{{%
    \normalfont B\kern-0.5em{\scshape i\kern-0.25em b}\kern-0.8em\TeX}}}

\setcopyright{acmcopyright}
\copyrightyear{2023}
\acmYear{2023}
\acmDOI{XXXXXXX.XXXXXXX}

\acmConference[ACM Trans. Comput. Healthcare] {ACM Transactions on Computing for Healthcare} {1} {1} 

\acmPrice{0}
\acmISBN{0}

\begin{document}



\title[Ubiquitous In-home Health Monitoring: A Comprehensive Survey]{Exploring the Landscape of Ubiquitous In-home Health Monitoring:\\A Comprehensive Survey}

\author{Farhad Pourpanah}
\email{f.pourpanahnavan@queensu.ca}
\orcid{0000-0002-7122-9975}
\author{Ali Etemad}
\email{ali.etemad@queensu.ca}
\orcid{0000-0001-7128-0220}
\affiliation{%
  \institution{Queens' University}
  \city{Kingston}
  \state{Ontario}
  \country{Canada}
  \postcode{K7L 3N6}
}

\begin{abstract}
Ubiquitous in-home health monitoring systems have become popular in recent years due to the rise of digital health technologies and the growing demand for remote health monitoring. These systems enable individuals to increase their independence by allowing them to monitor their health from the home and by allowing more control over their well-being. In this study, we perform a comprehensive survey on this topic by reviewing a large number of literature in the area. We investigate these systems from various aspects, namely sensing technologies, communication technologies, intelligent and computing systems, and application areas. Specifically, we provide an overview of in-home health monitoring systems and identify their main components. We then present each component and discuss its role within in-home health monitoring systems. In addition, we provide an overview of the practical use of ubiquitous technologies in the home for health monitoring. Finally, we identify the main challenges and limitations based on the existing literature and provide eight recommendations for potential future research directions toward the development of in-home health monitoring systems. We conclude that despite extensive research on various components needed for the development of effective in-home health monitoring systems, the development of effective in-home health monitoring systems still requires further investigation.  

\end{abstract}




\begin{teaserfigure}
\centering
  \includegraphics[width=0.9\textwidth]{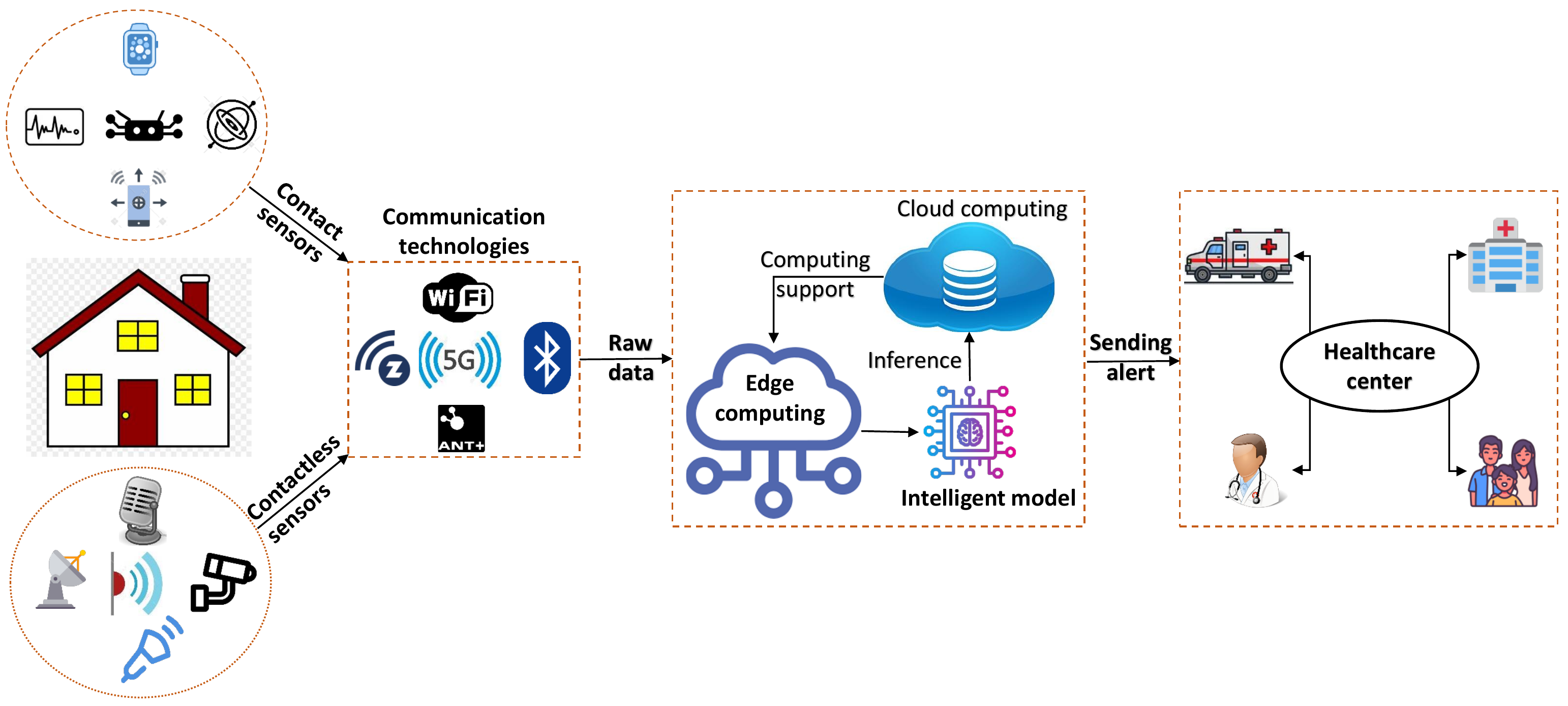}
  \caption{Overview of the ubiquitous in-home health monitoring systems.}
  \label{fig:over}
\end{teaserfigure}


\maketitle

\section{Introduction}
\label{sec:intro}
Continuous health monitoring is the process of regularly tracking and analyzing individuals' health data to identify/predict any possible health problems in their early stages. 
This is crucial for individuals, in particular older adults and people with disabilities, as it can provide valuable insights into their overall health status and help them diagnose/predict possible health issues~\cite{sarkar2021cardiogan}. Meanwhile, early detection of health problems through continuous monitoring allows individuals to take timely actions, reduce their stress, and gather valuable data through long-term observations~\cite{boulemtafes2016design}. In developing such systems, ubiquitous (or pervasive) technologies play a key role due to their ability to monitor various health parameters in real-time.

Ubiquitous computing refers to the idea of integrating computers into our surroundings. It is associated with the widespread integration of mobile/remote information and communication technologies that have some level of \textit{intelligence}, as well as network connectivity and advanced user interface, into our daily lives~\cite{weiser1999computer,orwat2008towards}. In health monitoring, this concept is widely used for collecting data, transmitting data, analyzing individuals' physical/mental and environmental conditions, and providing real-time feedback to assist individuals in managing their health. Accordingly, 
a robust ubiquitous healthcare system must satisfy six criteria, including 
(\textbf{1}) accessibility to multiple services from a healthcare provider,
(\textbf{2}) flexibility to adapt to changing patient needs and circumstances, 
(\textbf{3}) ensuring privacy during data exchange, 
(\textbf{4}) enabling remote data collection and monitoring, 
(\textbf{5}) providing personalized services, and 
(\textbf{6}) the capability of automatic decision-making and response~\cite{rodgers2014recent,ogunduyile2013ubiquitous}.

In recent years, due to several factors such as the aging population~\cite{kim2022home}, the impact of pandemics like COVID-19~\cite{pronovost2022remote}, and the need for personalized healthcare~\cite{wang2021multi}, there has been an increase in the development of in-home health monitoring systems to remotely monitor and track the health of individuals. This is particularly beneficial for those with chronic conditions, recovering from illness, and in need of ongoing medical attention. 
These systems aim to use ubiquitous technologies including mobile devices and applications, smart home devices, and wireless cloud/web services, to provide accessible, transparent, and reliable healthcare~\cite{ogunduyile2013ubiquitous,rodgers2014recent,fu2022short}. 
The ultimate goal is to monitor data in real-time and allow artificial intelligence (AI) systems or medical practitioners to remotely access individuals' health data, diagnose diseases, directly communicate with patients, and offer advice to them when needed~\cite{deen2015information}. 
In-home health monitoring systems have the potential to improve patient outcomes through continuous monitoring of chronic conditions, managing diseases, supporting the independence and safety of the aging population and individuals with disabilities, and enhancing their well-being and quality of life.
They can also increase patient satisfaction, ease the responsibilities of caregivers, and save costs by reducing hospitalization and other costly interventions~\cite{arar2021analysis,elouni2020intelligent,vandeweerd2020homesense}.

The field of in-home health monitoring is vast and complex, encompassing a variety of cutting-edge technologies, such as cloud computing, AI, and the Internet of Things (IoT)~\cite{puustjarvi2015role}. Furthermore, there remains a large gap between these advanced systems and their adoption in our homes~\cite{lussier2020integrating}. To obtain an inclusive understanding of these systems, in this survey, we aim to analyze the existing noteworthy systems in this area and answer the following questions.
\begin{enumerate}
    \item What is the current state of research on ubiquitous in-home health monitoring systems? Specifically, which sensing and communication technologies and algorithmic approaches have been utilized in the development of such systems?
    \item  What are the limitations and challenging issues of the existing ubiquitous in-home health monitoring systems?
    \item  What are the key future research directions that need to be further explored in order to address these issues?
\end{enumerate}

Our goal is to analyze the existing studies and identify different types of in-home health monitoring systems. We also explore various components of such systems and examine their applications. Additionally, we explore several interesting research directions that require further investigation. With this review, we hope to facilitate progress in developing novel and effective in-home health monitoring systems for improving the overall quality of life.

\subsection{Motivation and contributions} 
\label{SecSec:existing}
Several surveys on in-home health monitoring systems can be found in the literature~\cite{kim2022home,hobensack2022machine,alshamrani2022iot,fang2022smart,philip2021internet,fares2021directing,moyle2021effectiveness,nascimento2020sensors,wang2018leveraging,sujith2022systematic,tasoglu2022toilet}. 
Among them, the survey presented in~\cite{nascimento2020sensors} reviews sensors and systems developed for rehabilitation and health monitoring.
The survey papers in~\cite{hobensack2022machine,sujith2022systematic} center around the applications of machine learning (ML), deep learning (DL), and AI in-home healthcare, while, the works in~\cite{philip2021internet,alshamrani2022iot} focus on IoT-based in-home remote monitoring systems. Specifically, survey~\cite{philip2021internet} reviews IoT-based in-home remote monitoring systems and raises several concerns surrounding the reliability, privacy, security, and stability of such systems. On the other hand, the survey in~\cite{alshamrani2022iot} explores applications of health IoT (H-IoT) in the context of smart city infrastructure.
Lastly, in~\cite{fang2022smart}, four main factors in developing smart home care systems are identified. These factors are people, data, technology, and operational environments.

Besides, several literature focus on specific sub-areas or applications of ubiquitous in-home monitoring systems. For instance, the study in~\cite{tasoglu2022toilet} provides a comprehensive review of toilet-based in-home health monitoring platforms and discusses challenges and future perspectives. The review papers presented in~\cite{kim2022home,moyle2021effectiveness,fares2021directing,wang2018leveraging} mainly focus on older adults. For example, the work presented in~\cite{kim2022home} reviews the use of in-home monitoring technologies to monitor the well-being of older adults with a focus on monitoring daily activities, detecting abnormal behaviors, identifying cognitive impairment, detecting falls, indoor positioning, and evaluating sleep quality. 
In~\cite{moyle2021effectiveness}, the effectiveness of smart home technologies in assisting older adults with dementia is studied. The work in~\cite{fares2021directing} focuses on the utilization of information and communication technology solutions for aging people with chronic conditions with an emphasis on evaluating their impact on the quality of life from the biomedical point of view.
Moreover, the review paper in~\cite{wang2018leveraging} explores three works in the field of in-home health monitoring of older adults, highlights the gaps, and provides suggestions for developing practical applications such as diagnosis and nursing.

\begin{table*}[t]
\caption{Comparison between our survey paper and the existing ones on ubiquitous in-home health monitoring systems. Where ``\checkmark'' and ``{\huge +}'' indicate that the item is, respectively, fully and partially covered by the corresponding study, while ``$\times$'' indicates that the item is not surveyed.    }
\label{table:reviews}
\begin{adjustbox} {width=\columnwidth}
\begin{tabular}{lcccccccccccc}
\toprule
          & \multicolumn{12}{c}{Study} \\
                  \cmidrule{2-13}
 &   ~\cite{hobensack2022machine} & \cite{momin2022home}  &~\cite{fang2022smart} & ~\cite{fares2021directing} & \cite{philip2021internet}  &\cite{nascimento2020sensors} &~\cite{kim2022home}& ~\cite{alshamrani2022iot} &  \cite{wang2018leveraging} & \cite{sujith2022systematic} & \cite{tasoglu2022toilet} & Ours\\
\cmidrule{2-13}
    Sensing Technologies        & $\times$   & {\huge +}    & $\times$  & \checkmark& $\times$  &\checkmark &\checkmark & $\times$ &\checkmark& $\times$ &$\times$& \checkmark\\ 
    \midrule
    Communication technologies  & $\times$   & $\times$  & $\times$  & $\times$  & \checkmark& $\times$  & $\times$  & $\times$ & $\times$ & $\times$ &\checkmark& \checkmark\\
    \midrule              
    Intelligent \& computing systems   & {\huge +}      & {\huge +}           & \checkmark & {\huge +}         & $\times$   & $\times$  & $\times$  &\checkmark& $\times$ &  {\huge +}  &\checkmark&  \checkmark \\
    \midrule
    Applications                & $\times$  & {\huge +}       & $\times$   & \checkmark& $\times$  & $\times$  & \checkmark&\checkmark& $\times$ & $\times$ &$\times$& \checkmark\\
    \midrule
    Research gap             & \checkmark  & \checkmark & \checkmark & $\times$  &\checkmark &\checkmark & \checkmark&\checkmark&\checkmark&\checkmark&\checkmark&\checkmark \\
                  \bottomrule
\end{tabular}
\end{adjustbox}
\end{table*}

Our analysis of the above-mentioned survey papers indicates the lack of an in-depth review of ubiquitous in-home health monitoring systems that covers the various aspects and components of such systems under a unified framework.
Our goal in this survey is to answer the above-mentioned questions by exploring five key aspects of in-home health monitoring systems, including \textit{(1)} sensing technologies, \textit{(2)} communication technologies, \textit{(3)} intelligence and computing systems, \textit{(4)} application domains, and \textit{(5)} research gaps as well as exploring possible solutions to address these gaps. By analyzing these aspects, we intend to identify the advancements and limitations of in-home health monitoring systems. Table~\ref{table:reviews} compares our survey paper with prior works in this area. We observe that our paper provides a more complete, up-to-date, and comprehensive picture of the field.
In summary, the main contributions of our survey are as follows:
\begin{enumerate}
    \item We provide an overview of ubiquitous technologies used in-home health monitoring systems, identify their types, and discuss various components that are used to make up such technologies;
    \item We provide a review of sensing technologies, communication technologies, and intelligent and computing systems currently utilized in-home health monitoring systems;
    \item We provide a comprehensive overview of the diverse applications of ubiquitous technologies for in-home health monitoring systems;  
    \item We identify several significant challenges and research gaps in the development of in-home health monitoring systems and provide eight recommendations for future research direction.
\end{enumerate}

\subsection{Methodology}
\label{Sec:method}
\begin{figure}[t]
  \includegraphics[width=0.7\textwidth]{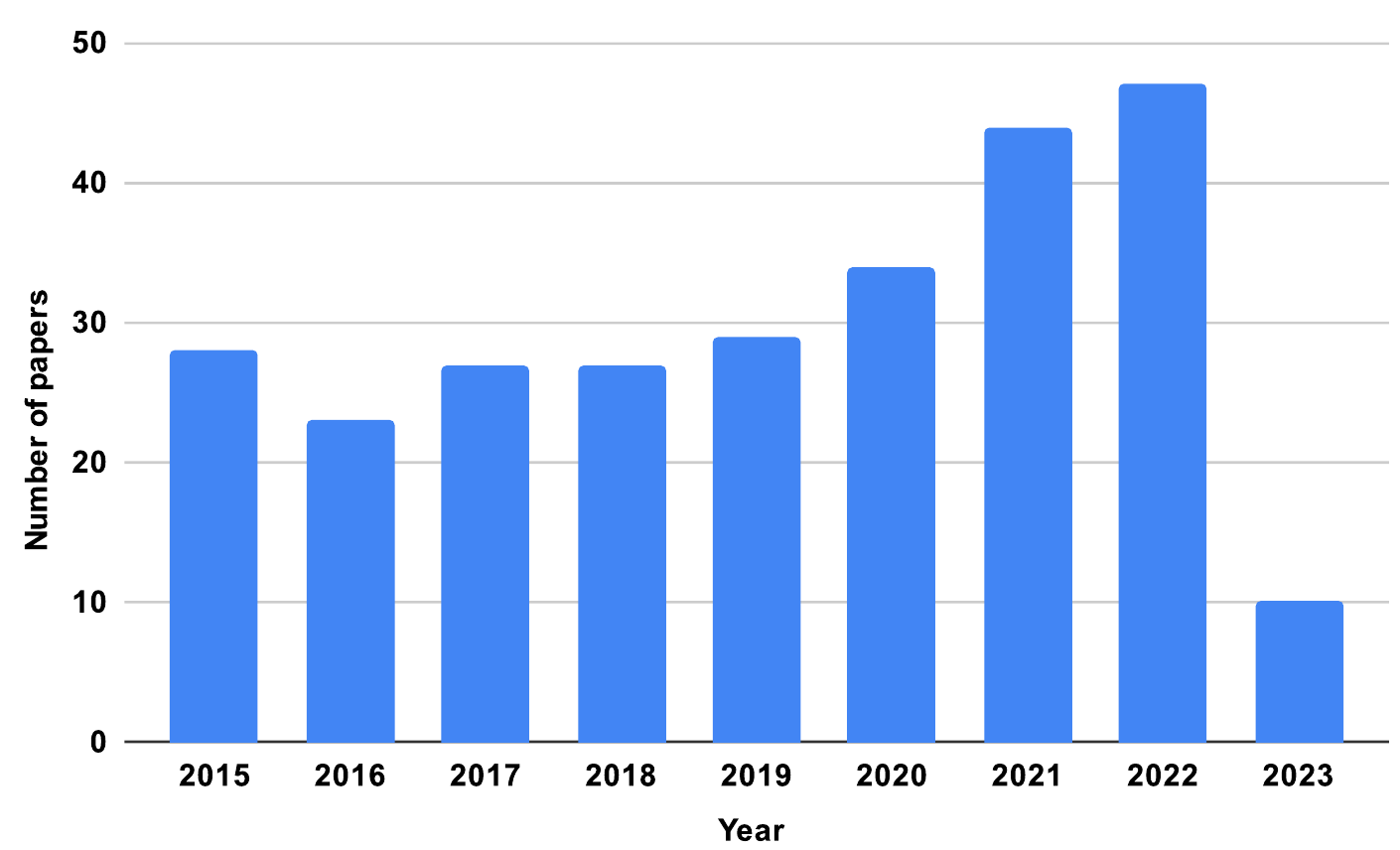}
  \caption{The publication trend of ubiquitous in-home health monitoring systems between 2015 and March 2023. The result is retrieved by searching five databases, including IEEE Xplore, ScienceDirect, ACM Digital Library, and Springer, as well as Google Scholar.}
  \Description{}
  \label{fig:docs}
\end{figure}

In this study, we review ubiquitous in-home health monitoring systems from 2015 onward. In April 2023, we searched four databases, namely IEEE Xplore, ScienceDirect, ACM Digital Library, and Springer, along with Google Scholar, for peer-reviewed articles written in English and related to the ubiquitous technologies for in-home health monitoring systems.
Specifically, we searched for ``in-home health monitoring'' OR ``monitoring health at home'' OR ``home health care'' AND ``ubiquitous technologies'' OR ``ubiquitous computing'' OR ``pervasive computing'' in the title, abstract, and keywords. We found a total of 269 papers published after 2015 in top-tier journals and conferences focusing on selected criteria. Fig.~\ref{fig:docs} shows the publication trend in the field, where the number of papers has steadily increased from 22 in 2016 to 48 in 2022. Among them, we identified 13 review papers, out of which 11 are presented in Subsection~\ref{SecSec:existing} (the other 2 were not quite relevant to this study). We then further studied the remaining 256 papers and removed 43 irrelevant ones. Additionally, we further excluded 12 papers related to healthcare that did not have a focus on in-home health monitoring, resulting in a set of 201 articles for further analysis.

\subsection{Organization}
\label{SecSec:org}

This paper consists of a total of five sections including the Introduction. Next, Section~\ref{Sec:overview} provides an overview of in-home health monitoring systems. Specifically, we discuss various types of in-home health monitoring systems and identify their main components, including sensing technologies, communication technologies, cloud computing, security and data governance, and intelligence and computing systems. We also provide a detailed discussion of each component. Section~\ref{Sec:APP} reviews application domains of in-home health monitoring systems. In Section~\ref{Sec:discussion}, we provide discussions and identify existing research gaps in the field, and outline future research directions. 
Finally, we conclude our survey in Section~\ref{Sec: con}.

\section{Components and Technologies} 
\label{Sec:overview}
Over the years, there has been significant progress in the development of in-home health monitoring systems, resulting in various form factors and designs. In general, these systems consist of several components and technologies (see Fig.\ref{fig:over}), including sensing technologies, communication systems, and intelligent and computing systems~\cite{moreira2019intelligent,ghamari2016survey}, as follows:

\begin{itemize}
    \item \textbf{Sensing technologies} include all sensors that are used to collect individuals' health-related data, e.g., vital signs, environmental data, and others. We provide a review of these technologies in Section~\ref{Sec:sensors}.    
    \item \textbf{Communication technologies} include all technologies that facilitate data transmission between various sensors, devices, and locations. Section~\ref{SecSec:IoT} provides a brief review of communication technologies commonly utilized in ubiquitous in-home health monitoring systems. 
    \item \textbf{Cloud computing} includes systems that enable the storage and processing of large amounts of data collected by various sensors and devices in the home. We discuss these systems in detail in Section~\ref{Sec:cloud}. 
    \item \textbf{Security and data governance} involve making sure data is handled safely. This protects patient information and prevents unauthorized access. We discuss the security and data governance in the area of in-home health monitoring systems in Section \ref{secsec:sec}.
    \item \textbf{Machine learning and deep learning techniques} include all intelligent models/algorithms that are employed to analyze data for extracting insights, providing feedback, or making decisions. These techniques are summarized in Section~\ref{sec:ML}.
    
\end{itemize}

\subsection{Sensing technologies} 
\label{Sec:sensors}
Data plays a vital role in ubiquitous in-home health monitoring systems~\cite{puustjarvi2015role}, which can be obtained from various sensing technologies. In this survey, we broadly categorize these technologies into two groups: contact and contactless. The first group includes all sensors that detect or measure parameters by physical contact with a body or surface. Contact sensors can be embedded into wearable devices, e.g., smart watches, wristbands, patches, and others, for monitoring/tracking various vital signs such as heart rate (HR), blood pressure (BP), body temperature (BT), and sleep patterns. 
Additionally, these devices can monitor personalized medical conditions like diabetes~\cite{jing2017intelligent} by measuring and tracking blood glucose and other chemical compounds.
In contrast, contactless sensors include various sensing devices installed throughout the home to acquire data without requiring any contact with the body or surface. These sensors include cameras, microphones, and ambient sensors to acquire environmental data, e.g., humidity room temperature, and others. 
 
\begin{figure}[t]
  \includegraphics[width=0.9\textwidth]{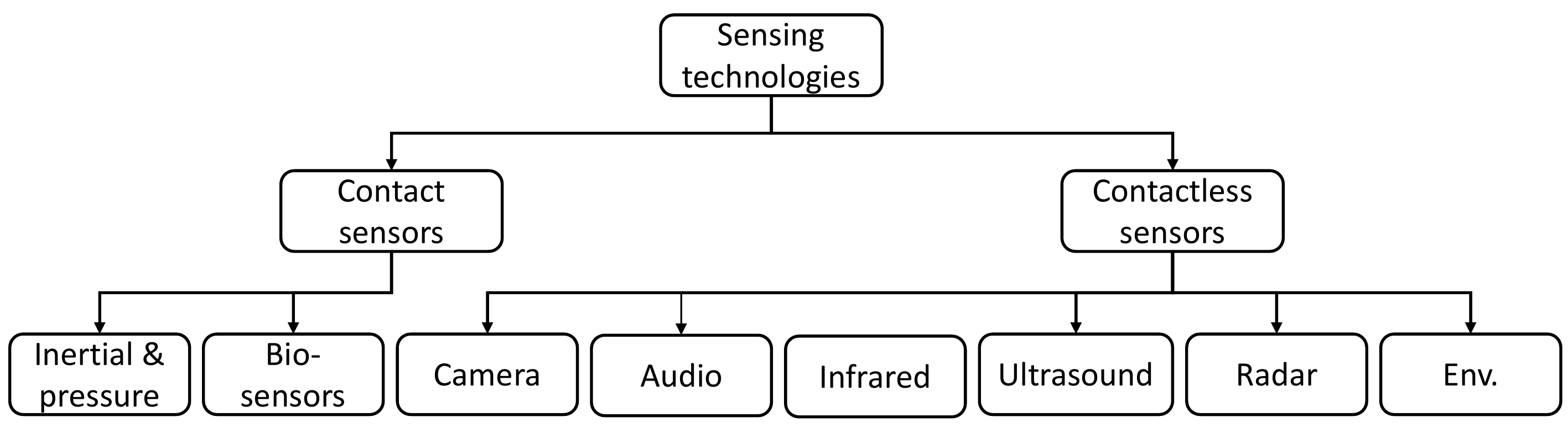}
  \caption{Categorization of sensing technologies for data acquisition for in-home health monitoring systems. We categorized these technologies into two sub-groups, namely contact and contactless sensors, and then identified various sensors for fell under each of these groups.}
  \Description{}
  \label{fig:sensors}
\end{figure}

In the following sections, we present a hierarchical categorization of each group (see Fig.~\ref{fig:sensors}) and discuss how each category works. 
Additionally, we refer readers interested in sensing technologies to survey papers~\cite{wang2021unobtrusive,singh2020sensor}. In~\cite{wang2021unobtrusive}, a review of sensing technologies for unobtrusive in-home health monitoring is presented, while~\cite{singh2020sensor} provides a survey on sensing technologies specifically developed for fall detection~\cite{singh2020sensor}. 

\subsubsection{Contact sensors}
We classify this category of sensors into two groups: (1) inertial and pressure sensors, and (2) bio-sensors. The inertial and pressure sensors can be further divided into the following: 

\begin{itemize}
    \item \textbf{Inertial measurement units (IMUs)} consist of accelerometers, gyroscopes, and magnetometers that are used to monitor individuals' activities in ubiquitous in-home health monitoring systems~\cite{singh2020sensor} such as wearable devices. Accelerometers measure acceleration and are the most common type of sensor utilized to monitor vibration, acceleration, and movement. 
    Gyroscopes measure angular velocity and allow for much more nuanced motion analysis alongside accelerometers. Finally,  
    magnetometers measure the changes in the magnetic field, which can be used for better detection of orientation.  
    Together, IMU sensors are often embedded in wearable devices such as smartwatches and wristbands to detect gait parameters~\cite{hahm2022home}, measure distance and the number of steps, calorie expenditure~\cite{candy2021accelerometer}, classify the type of activity \cite{rahimi2021self}, and others.
   
    \item \textbf{Plantar stress distribution (PSD) sensors} are used for monitoring foot function during standing and walking activities. They compute the amount of pressure distribution on the soles of feet during standing/walking~\cite{zhao2021flexible}.

    \item \textbf{Floor pressure sensors} measure the force exerted on the surface as a result of the mass and impact of the subject. These sensors can be used to identify and locate objects, detect falls~\cite{muheidat2020home}, monitor balance~\cite{javaid2017balance} and asses gait~\cite{cantoral2015intelligent}, and others.   
\end{itemize}

Bio-sensors include those used to identify and determine biological signals in health monitoring systems. These sensors include the following:

\begin{itemize}
    \item \textbf{Electrocardiogram (ECG)} provides information related to the heart by obtaining the relevant electrical signals and converting them into a format that is understandable by clinicians~\cite{dey2017developing}. These sensors are commonly used to detect arrhythmia~\cite{soltanieh2022analysis}, assess stress~\cite{behinaein2021transformer}, cardiac monitoring~\cite{wang2023ecg}, atrial fibrillation detection~\cite{jalali2019atrial}, and even recognize emotions~\cite{sarkar2020self}, along with other heart-related issues.    
    
    \item\textbf{Photoplethysmogram (PPG)} and \textbf{blood volume pulse (BVP)} are used to measure changes in blood volume. 
    PPG sensors measure specific wavelengths of light reflected back from tissues to estimate volumetric changes in blood flow.  
    In contrast, BVP sensors use a mechanical transducer such as piezoelectric crystals to measure blood volume changes. These sensors are used to measure several parameters such as HR~\cite{biswas2019heart}, BP~\cite{el2020review}, oxygen levels~\cite{lee2023fiber}, and others. 
   
    \item \textbf{Pulse oximeters} are employed to determine the oxygen saturation levels (SpO2) in the blood. These sensors emit two wavelengths of light into the skin and measure the amount of absorbed light by oxygen-carrying hemoglobin molecules in the blood. Monitoring the SpO2 is important for patients with chronic diseases after surgery~\cite{watthanawisuth2010wireless}.
     
    \item \textbf{Electroencephalography (EEG)} records electrical activity in the brain by placing electrodes on the scalp. EEG is an effective sensor for monitoring brain activity both during waking and sleeping states. These sensors are used to classify intended actions~\cite{zhang2019classification}, recognize emotions~\cite{zhang2022parse,zhang2021deep}, assess stress~\cite{sharma2022evolutionary}, estimate cognitive load~\cite{angkan2023multimodal}, and others.   
   
    \item \textbf{Electromyography (EMG)} provides information related to muscles by measuring the electrical activity or muscle response when stimulated by a nerve~\cite{nascimento2020sensors}. EMG sensors can be used to identify neuromuscular abnormalities~\cite{fidanci2023needle}, estimate the force generated by muscles during different types of movements~\cite{hajian2022bagged,hajian2022multimodal,hajian2020automated}, and others.  

    \item \textbf{Electrodermal activity (EDA)}/\textbf{galvanic skin response (GSR)} measures the electrical conductivity of the skin, which is indicative of the activity of sweat glands and the sympathetic nervous system. These sensors are used to detect seizures~\cite{poh2012convulsive}, measure stress levels~\cite{liu2018psychological}, emotion recognition~\cite{bhatti2022attx}, and other related issues. 
   
    \item \textbf{Bioimpedance (BIA)} measures the impedance of biological tissues in the body by sending a small electrical current through the body and computing the resistance of the tissues to the flow of this current. The measured impedance can be used to estimate various body composition parameters, such as the percentage of body fat~\cite{silveira2020body}, muscle mass~\cite{cheng2021diagnosis}, and others.  
   
    \item \textbf{Temperature} sensors are used to measure body temperature. Thermometers are the most commonly used contact-based sensors placed under the tongue or armpit to measure body temperature, which can be used to detect fevers~\cite{am2021survey,al2021design}.   
   
    \item \textbf{Airflow} measures the flow of air in and out of the respiratory system during breathing to monitor respiratory function~\cite{moshizi2021polymeric} and detect abnormalities such as sleep apnea~\cite{lin2016sleepsense}. 

\end{itemize}

\begin{figure}[t]
  \includegraphics[width=0.6\textwidth]{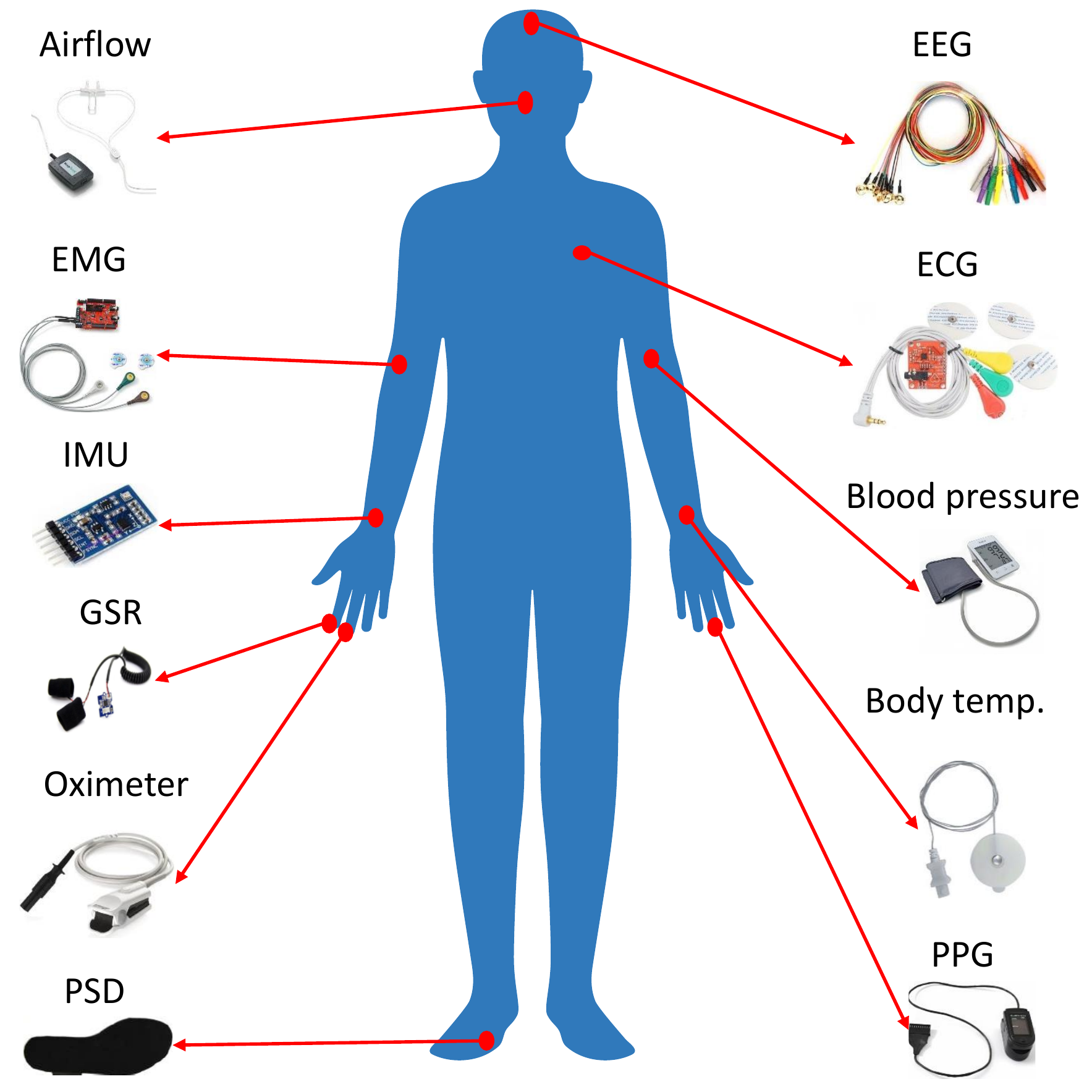}
  \caption{ Various sensors used for data acquisition from the human body.}
  \Description{}
  \label{fig:body}
\end{figure}

Fig.~\ref{fig:body} provides a comprehensive illustration of the different types of sensors that can be employed for data acquisition from the human body for the purpose of in-home health monitoring. This data can include HR, muscle activity, brain waves, BT, and more, enabling accurate analysis and interpretation of the subject's physiological state.

\subsubsection{Contactless sensors}
This category of sensors offers a non-invasive approach for detecting and monitoring various types of data for in-home health monitoring, as they do not need to maintain contact with the body. These sensors include the following:  

\begin{itemize}

    \item \textbf{Camera} sensors can be placed on tables, cabinets, or installed on the walls or ceilings, to capture visual health-related information. These sensors can be used to identify and track individuals~\cite{chiang2018kinect}, estimate pose~\cite{chen2021wide}, detect falls~\cite{luo2018computer}, and even detect HR through remote PPG estimation~\cite{liu2021motion}, among others.

    \item \textbf{Audio} sensors are used to detect and analyze sounds, such as snoring~\cite{xie2023assessment} and coughing patterns~\cite{sharan2022automated}, which can be used to measure various factors related to the health of individuals.
   
    \item \textbf{Infrared (IR)} sensors operate based on infrared radiations, which can be used to measure temperature~\cite{hildebrandt2010overview} and analyze motion~\cite{wang2021multi}.
   
    \item \textbf{Ultrasound} sensors leverage the advantage of high-frequency sound waves to determine the distance and proximity to an object. In ubiquitous in-home health monitoring systems, these sensors are employed to detect several health-related parameters such as HR and breathing~\cite{koval2016distance}.
   
    \item \textbf{Radar} sensors utilize radio waves to identify motion and measure vital signs. In recent years, the use of radar technology for indoor human monitoring has seen a rise in applications such as gait assessment~\cite{abedi2023ai,seifert2019toward}. This considerable attention is mainly due to its reliability, safety, privacy-preserving, and ability to effectively monitor subjects~\cite{seifert2019toward}.

    \item \textbf{Environmental} sensors measure environmental parameters such as air quality~\cite{moore2018managing}, humidity~\cite{honda2022wearable}, and temperature~\cite{siam2022portable}, which can be applied for in in-home health monitoring systems.
    
\end{itemize}

\subsubsection{Discussion on contact vs. contactless sensors}
According to the discussion provided in the previous subsections, it is evident that contact and contactless sensors serve distinct purposes, each with its strengths and weaknesses. 
Given the direct contact with the body and hence more proximity to users in comparison to contactless sensors, contact sensors can often monitor more intricate information over more sustained periods.  
However, it may not always be possible or practical to have subjects wear sensors consistently and reliably. Moreover, personal hygiene is another concern with contact sensors. The general requirement to design and develop such devices in small wearable or mobile form factors may also result in a trade-off with respect to acquisition quality. Additionally, such devices often require constant charging and wireless access to the Internet.
In contrast, contactless sensors play an essential role in situations where physical distancing is crucial. They provide solutions that are not feasible with contact sensors. Their ability to perform indirect measurements makes them especially effective in sensitive environments. However, this benefit comes with certain challenges. For instance, the data collected by contactless sensors is often less ubiquitous than wearables given the need for the subject to be physically in the presence of the sensor. Additionally, distinguishing between the data collected from subjects can be challenging in comparison to contact sensors which are often considered personal devices. 
Therefore, choosing between contact and contactless sensors depends on the specific requirements of the application, the environment, the type of data that needs to be analyzed, and the ability and willingness of the subjects to wear/use sensing devices.

\subsection{Communication technologies}
\label{SecSec:IoT}
The Internet of Things (IoT) is a collection of interconnected sensors, devices, and systems that communicate with each other through the Internet~\cite{yuehong2016internet,alshamrani2022iot}. IoT is a rapidly growing technology that offers plenty of opportunities to improve in-home health monitoring~\cite{farooq2022blockchain,rajasekaran2020health,zhang2020sparse}. Integrating IoT into in-home health monitoring systems, which is also known as the Internet of medical things (IoMT)~\cite{alshamrani2022iot} and health industrial Internet of things (HealthIIoT)~\cite{hossain2016cloud}, leads to a robust system that empowers individuals to be more involved in their healthcare and enables healthcare providers to effectively monitor patients' health status. While IoT technologies have the potential to revolutionize in-home health monitoring, challenges such as privacy and data security remain open problems in the area.

IoT technologies rely on various communication technologies to transmit data between sensors, devices, and intelligent systems. These communication technologies can be classified as wired and wireless. Wired communication technologies utilize either local area networks~\cite{clark1978introduction} or wide area networks~\cite{raza2017low} to connect devices and transmit data. Moreover, Ethernet~\cite{winzer2010beyond} and Powerline communication (PLC)~\cite{majumder2004power} are two widely used wired communication technologies. Ethernet is a standard wired communication system that uses a protocol to establish connections between devices, while PLC enables communication between devices through power lines.

In contrast, wireless communication technologies use various types of wireless protocols to transfer data between devices, which can be categorized into \cite{wang2022evolution,mahmood2015review}: 
\begin{itemize}
    \item \textbf{Short-range wireless communications \cite{wang2022evolution}} entail a variety of wireless technologies designed to operate over small distances, such as Bluetooth, Zigbee, Z-Wave, WiFi, and ANT+. Typically, these technologies facilitate the creation and maintenance of personal area networks. They enable the interconnection of devices within a close range, usually not extending beyond a few meters. The primary advantages of short-range wireless communication include its low power consumption and minimal infrastructure requirements. In the context of in-home health monitoring systems, these technologies enable the connection of various health monitoring devices and sensors within a home environment. For instance, wearable devices like HR monitors, BP cuffs, and glucose monitors can wirelessly transmit vital health data to a central unit. This setup allows continuous monitoring of a patient's health status without impeding their daily activities. 
  
    \item \textbf{Local area wireless communications \cite{uwaechia2020comprehensive}} provide wireless connectivity over relatively large areas such as homes, offices, and campuses. This category includes technologies like WiFi, which is ubiquitous in its application for creating wireless local area networks (WLANs). Their capability to deliver high-speed data transfer rates makes them ideal for a wide range of uses. 
    Furthermore, advancements in this field, like the development of WiFi 6, continue to push the boundaries in terms of speed, capacity, and efficiency \cite{oughton2021revisiting}. They accommodate a large number of connected devices and increase data-intensive applications. In the context of in-home health monitoring systems, they offer substantial benefits. For instance, WiFi can serve as the backbone for connecting various health-related devices and sensors throughout a home. This network allows for the continuous transmission of health data from different parts of the residence to a centralized system, where it can be accessed by healthcare providers, caregivers, or AI systems. For patients with chronic illnesses or elderly care, devices like BP monitors, emergency response systems, and motion sensors can be integrated into this network. The reliable and constant connectivity ensures real-time monitoring and immediate data analysis, facilitating prompt medical responses and enabling effective remote patient monitoring \cite{pattnaik2022future}. 
    
    \item \textbf{Wide area wireless communications \cite{ikpehai2018low}} include a range of technologies designed to provide network coverage over large geographical areas, far surpassing the capabilities of local area wireless communications. This category is predominantly defined by cellular networks, such as 3G, 4G, and the latest 5G technologies \cite{khan2019survey}. They facilitate mobile communication and data services over extensive areas. A distinguishing feature of this category is its ability to maintain consistent communication links in diverse environments, ranging from densely populated urban areas to remote rural locations. The evolution from 3G to 4G brought significant enhancements in terms of data transfer speeds and reliability, while the advent of 5G is set to revolutionize the landscape with even faster speeds, lower latency, and the capacity to connect a massive number of devices simultaneously. In the context of in-home health monitoring systems, wide area wireless communications play a transformative role. The robust and high-speed connectivity offered by these networks enables the transmission of health data from in-home monitoring devices directly to healthcare providers, regardless of distance. This aspect is particularly vital for remote patient monitoring, where real-time data transmission can facilitate prompt medical responses and interventions. For instance, patients with chronic conditions can have their health parameters like HR, BP, and glucose levels continuously monitored and sent to their doctors for immediate analysis. This ongoing monitoring can lead to timely adjustments in treatment plans, potentially preventing hospital readmissions. Furthermore, the reliability and wide coverage of these networks ensure that even patients in remote areas can access quality healthcare services \cite{raza2017low}. 

\end{itemize}

Wired communication technologies, known for their fast speeds, security, and reliability, are particularly practical in monitoring systems where devices and sensors are close. However, their use is constrained by reduced flexibility, higher costs, and the need for continuous maintenance. On the other hand, wireless communication systems, embedded in contact-based sensing technologies like mobile or wearable devices, stand out for their flexibility and scalability. Despite these advantages, wireless systems often face challenges in security and are susceptible to signal interference from other devices. The studies in~\cite{ahmed2022wireless} and~\cite{am2021survey} provide extensive surveys on wireless sensor networks and their data transmission capabilities.

\subsection{Cloud computing} 
\label{Sec:cloud}
As stated earlier, ubiquitous in-home health monitoring systems could involve many interconnected devices that acquire large amounts of data. To effectively store and process such data, cloud computing offers an elegant and scalable solution~\cite{hashem2015rise}. It is pivotal in providing the necessary resources and infrastructure required for storage and complex computations that ML and DL models often demand.
While cloud computing platforms are widely used in a variety of applications, they generally introduce delays in processing. Moreover, they often deal with security concerns due to sharing data with third-party organizations for data management~\cite{tasoglu2022toilet}. To alleviate these issues, edge computing~\cite{shi2016edge} and fog computing~\cite{bonomi2012fog} have emerged as viable solutions. Edge computing performs data processing and analysis at the edge of the network, where the term `edge' refers to computing and network resources in between the sensing technologies and the cloud centers. The rationale behind edge computing is to perform computation close to the data source so that there are fewer opportunities for the data to become exposed to security risks. Similar to edge computing, fog computing distributes data processing and analysis across multiple nodes in the network, such as edge devices, network routers, and cloud servers. The difference between edge computing and fog computing is that edge computing is focused on processing data at the edge of the network where the data is generated, while fog computing distributes processing across multiple nodes in the network~\cite{shi2016edge}. Both edge and fog computing can reduce the response time and preserve the privacy and security of the data. However, these frameworks often suffer from limited storage, network bandwidth, and processing power when compared with cloud computing.

\subsection{Security and data governance}
\label{secsec:sec}

Security and strong data governance measures are essential in the development of in-home health monitoring systems \cite{kumar2023establishment,gharaibeh2017smart} to prevent various attacks and security threats, and preserve privacy \cite{philip2021internet}, particularly as these systems may store and share sensitive data with third parties \cite{newaz2021survey,miao2022real,srivastava2018automated,hadjixenophontos2023prism}.
Accordingly, various studies tackle data security for health monitoring. As stated in Section \ref{Sec:cloud}, cloud computing platforms usually suffer from security issues, while edge and fog computing have proven to be promising alternative solutions \cite{tasoglu2022toilet}. The study in \cite{miao2022real} describes a matrix encryption method paired with ML for secure biomedical signal processing. The work in \cite{meingast2006security} investigates the security consequences of healthcare monitoring systems. The study in \cite{harvey2020security} suggests a framework for smart medical device security in home environments. The study in \cite{tran2023vpass} introduces a framework for managing personalized privacy requirements in voice assistant systems. The importance of security in these systems is further elaborated in \cite{somaya2019secure,alami2018study}, and a comprehensive review of potential attacks and their impacts is provided in \cite{newaz2021survey}. The works in \cite{gupta2023privacy,bian2023verifiable} propose privacy-preserving approaches for remote HR monitoring. 
 On the other hand, to enhance the privacy of data, frameworks such as federated learning \cite{mcmahan2017communication} and Blockchain~\cite{aujla2020decoupled,jita2018framework}  can be adopted. These techniques distribute the data across the network rather than using a single unit to control the whole data. Federated learning is a privacy-preserving ML paradigm that allows multiple clients, such as homes, to collaborate in training a model without sharing their data. By using federated learning, data can be collected from different houses and sources to develop a robust in-home health monitoring system.
Blockchain~\cite{kube2018daniel} is a disruptive technology that enables secure transactions between individuals in unreliable environments without the need for a central party. Recent studies have adopted federated learning \cite{chen2021establishing,wu2020fedhome}, Blockchain ~\cite{aujla2020decoupled,jita2018framework}, and both~\cite{farooq2022blockchain} in developing in-home health monitoring systems. 

\subsection{Machine learning and deep learning techniques}
\label{sec:ML}
ML is a sub-field of AI that uses data and complex algorithms to learn from data and make predictions~\cite{hassanien2019machine,bishop2006machine}. Supervised ML techniques were used as the gold standard in early in-home health monitoring systems. The most popular classic ML techniques use for health applications include linear regression (LR)~\cite{stapleton2009linear}, decision tree (DT)~\cite{kotsiantis2013decision}, random forest (RF)~\cite{breiman2001random}, support vector machine (SVM)~\cite{noble2006support}, multilayer perceptron (MLP)~\cite{gardner1998artificial}, and fuzzy models~\cite{wang2021fuzzy}. Some popular applications of ML techniques for in-home health monitoring systems include monitoring daily activities~\cite{amin2016radar,pham2018delivering}, fall detection (FD)~\cite{castillo2022low,vishnu2021human}, sleep monitoring~\cite{lin2016sleepsense,hsu2017zero}, disease detection~\cite{hossain2016cloud,dharmik2021iot} and monitoring~\cite{miao2018wearable,chatterjee2018designing}, etc. These applications, among others, are listed in Tables~\ref{table:DA} to \ref{table:mhm}.

Despite their great progress, classic ML algorithms extensively rely on engineered features to generate results. Moreover, recent advances in digital technologies have resulted in large datasets collected from various sensing technologies which traditional ML algorithms are not able to cope with efficiently and effectively. 
 To address these issues, DL, a subset of machine learning, consists of techniques that are effective in extracting representations from large amounts of raw data, without the need for human feature engineering~\cite{yan2015deep,mahmud2018applications,wang2020recent,heidari2023hiformer,pourpanah2022review}.
These methods use multiple layers of linear and non-linear transformations to learn the relations between inputs and their corresponding outputs. 
The main types of commonly used DL architectures are convolutional neural networks (CNNs)~\cite{wiatowski2018mathematical}, recurrent neural networks (RNNs)~\cite{pascanu2013construct}, and transformers~\cite{vaswani2017attention}.

CNNs are a type of DL network that is inspired by the visual cortex of animals~\cite{hubel1968receptive}. Typically, CNNs consist of three types of layers: convolutional, pooling, and fully connected layers. The convolutional and pooling layers are usually located in the lower levels of the network and are responsible for extracting relevant features from the input samples. These layers apply a set of filters to the input data, generating a set of linear activations that are then passed through non-linear functions~\cite{hinton2012improving}. This helps to reduce the complexity of the input data. Following the convolutional layers, pooling layers are used for down-sampling the filtered results. These pooling layers reduce the size of the activation maps by transforming them into smaller matrices~\cite{schere2010evaluation}. This process helps to mitigate the overfitting problem by reducing the complexity of the data~\cite{ngyen2015deep}. The fully connected layers are typically located after the convolutional and pooling layers, and their purpose is to learn more abstract representations of the input data. In the last layer of the CNN, a loss function, such as a softmax classifier, is used to map the input data to its corresponding output.
CNNs have demonstrated remarkable performance on various health monitoring tasks, including but not limited to, activity monitoring~\cite{abedi2023ai}, bio-signal analysis~\cite{sarkar2020self}, affective computing~\cite{sepas2019deep}, medical image analysis \cite{wang2023attentive,wang2023dc}, and others.
The main strength of these models is their ability to learn hierarchical representations from input data, which allows them to handle variations such as changes in ambient conditions, scale, and orientation. However, CNNs demand a large number of training data and high-performance computing resources, and their predictions are often challenging to interpret~\cite{rawat2017deep}.

RNNs are a class of DL architectures that are designed to model/process sequential data, such as natural language and time series~\cite{rumelhart1986learning}. The nodes in each layer of RNNs are connected, which enables RNNs to handle sequential data. RNNs learn a representation of the sequential data and maintain a hidden state that captures the context of the sequence up to the current time step. They produce outputs based on the hidden state and the input at the current time step. 
Despite the success of RNNs in handling sequential data, remembering/learning from past data, and flexibility in a range of tasks, they are computationally expensive and face difficulty in handling variable-length inputs~\cite{yu2019review}. Moreover, they are difficult to train on long sequences of data due to vanishing or exploding gradient problems~\cite{dang2017asurvey}. To alleviate the vanishing gradient problem, improved versions of RNNs, such as long short-term memory (LSTM)~\cite{hochreiter1997long} and gated recurrent units (GRU)~\cite{chung2014empirical}, use gate units to decide what information to keep or remove from the previous state. These networks have been widely used for fall detection and modeling health-related signals such as bio-signals~\cite{zhang2019classification,xie2022passive}.

Recently, transformers have shown promising results in the field of natural language processing~\cite{lin2022survey,vaswani2017attention} and computer vision~\cite{dosovitskiy2020image,touvron2021training}. They are based on the idea of self-attention, where the model learns to attend to different parts of the input sequence to make predictions. They first transfer the input sequence into a sequence of embeddings by processing it through a series of self-attention and feedforward layers. The self-attention mechanism enables the model to selectively attend to various parts of the input sequence, while the feedforward layers assist in capturing complex interactions between the input tokens~\cite{lin2022survey,jaderberg2015spatial}.  
Unlike RNNs that sequentially process the data, transformers can process data in parallel. Thus they are faster than RNNs. However, they require massive amounts of data for successful training. These models have recently begun to be explored in the context of ubiquitous health monitoring, for instance in arrhythmia detection~\cite{hu2022transformer}, stress detection~\cite{behinaein2021transformer}, and more.

Autoencoders are a type of neural network architecture that can be used for representation learning. They consist of two components: (1) an encoder, and (2) a decoder. They use the encoder to learn a compressed or encoded representation of the input data, and then the decoder is used to reconstruct the learned representation back into the original data~\cite{hinton2006reducing,mohamud2023encoder}. One of the key advantages of Autoencoders is their ability to learn useful representations from input data without the need for labeled data. This learning approach makes them effective for a range of applications such as dimensionality reduction, anomaly detection, and data compression. More importantly, Autoencoders are flexible in terms of the types of data they can handle. The encoder and decoder components can be implemented using different types of neural networks, including MLPs, CNNs, or RNNs, which enables them to handle both static and sequential data. Autoencoder architectures have been used for several different health-related applications such as affective computing~\cite{ross2021unsupervised}, arrhythmia classification~\cite{hou2019lstm}, and heart disease diagnosis~\cite{deperlioglu2020diagnosis}.

Generative models learn to synthesize new data based on existing data, which can be used to improve the training of DL models~\cite{denton2015deep} and address imbalanced classification problems by generating new samples for underrepresented classes~\cite{guo2004learning,jalali2023adversarial}. There are a variety of different types of generative methods including generative adversarial networks (GANs)~\cite{goodfellow2020generative}, Variational Autoencoders (VAEs)~\cite{kingma2022autoencoding}, diffusion models~\cite{ratcliff1978theory}, normalizing flows~\cite{rezende2015variational}, and autoregressive models~\cite{oord2016wavenet}, among others. Among the more popular approaches to generative models, GANs consist of a generator and a discriminator. The generator takes a random noise vector as input and produces a new sample that mimics the training data, while the discriminator aims to distinguish real samples from the generated ones. On the other hand, VAEs consist of an encoder that maps input data onto a latent space and a decoder that maps the latent space to the sample. These models have been used to 
translate PPG to ECG signals for cardiac monitoring \cite{sarkar2021cardiogan} address the imbalanced and non-IID distribution problem in federated earning~\cite{wu2020fedhome}, and more.  

DL models can be trained using supervised and unsupervised learning. Supervised models are trained based on labeled samples, where each training sample is associated with an output or label. These models learn a mapping function between input and output by minimizing an objective function and have been widely used for in-home health monitoring~\cite{liu2020privacy,luo2018computer,siam2022portable}. In contrast, unsupervised models are trained without labeled data. They aim to learn representations that are generalizable to various downstream learning problems. In other words, unsupervised models explore meaningful relationships such as clusters or groups of samples that contain the same structure/information within the data. 

 \begin{figure}[tb]
  \includegraphics[width=0.9\textwidth]{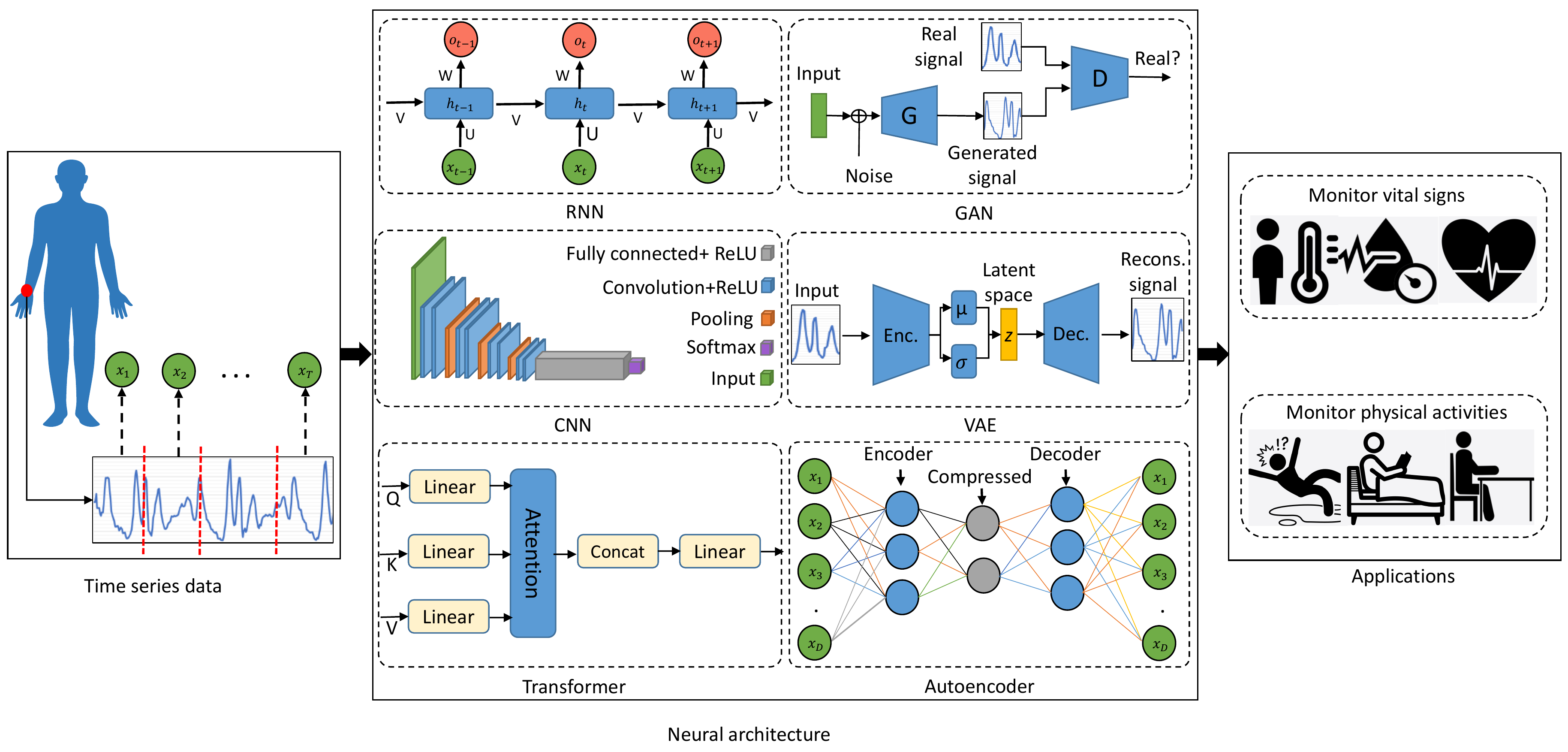}
  \caption{ An illustration depicting the employment of various DL techniques for the processing of sequential time-series data. The model inputs raw signal data at different time steps $(x_1, x_2, ..., x_T)$ and analyzes patterns using DL techniques to assess overall wellness.}
  \Description{}
  \label{fig:FDRNN}
\end{figure}

Semi-supervised~\cite{van2020survey} and self-supervised~\cite{jing2020self} approaches have recently shown impressive performances in learning generalizable representations from data, which have also been used in the field of health monitoring~\cite{sarkar2020self,zhang2022parse,sarkar2021detection}. Semi-supervised models learn from a large number of unlabeled samples and a small set of labeled samples~\cite{gu2020asemi,pourpanah2023ensemble}. 
Self-supervised models, on the other hand, learn from unlabeled data on which specific augmentations or transformations have been applied. They extract useful representations from the data by solving pre-text tasks such as predicting the rotation angle of an image or solving a jigsaw puzzle. The learned model can be fine-tuned to solve downstream tasks such as classification and segmentation problems. Given the general difficulty of recording labels for ubiquitous health data, the application of semi- and self-supervised methods in this area has begun to result in significant contributions. Examples include ECG-based affect~\cite{soltanieh2023distribution,sarkar2020self} and healthcare~\cite{krishnan2022self,huang2023self,dave2023timebalance}.

Fig.~\ref{fig:FDRNN} provides an overview of different DL models often used for in-home health monitoring applications.  Time-series data is segmented into sequential inputs $(x_1, x_2, ..., x_T)$ that are fed into a DL model before further processing. The selection of the DL model often relies on the particular application. For instance, CNNs are notable for hierarchical feature learning, crucial in image-based health monitoring \cite{chen2021establishing}, yet they require extensive data and computational power and suffer from interpretability issues \cite{zhang2021survey}. RNNs have been widely used to handle sequential data \cite{yu2019review}, but struggle with long sequences, although LSTM and GRU variants offer some solutions \cite{yang2020lstm}. Transformers have recently shown promise in processing both images \cite{khan2022transformers} and sequences \cite{wen2022transformers} via self-attention, but they also demand large training datasets \cite{zaheer2020big}. Autoencoders are used for unsupervised representation learning \cite{tschannen2018recent} and have been shown to be effective in applications like anomaly detection \cite{chen2018autoencoder}. Generative models like GANs and VAEs are instrumental in synthesizing realistic data \cite{bond2021deep}, aiding in imbalanced datasets and training robustness \cite{harshvardhan2020comprehensive}, but they can be complex to train and require careful tuning.

\vspace{4mm}

\section{Applications}
\label{Sec:APP}
\begin{figure}[t]
  \includegraphics[width=0.9\textwidth]{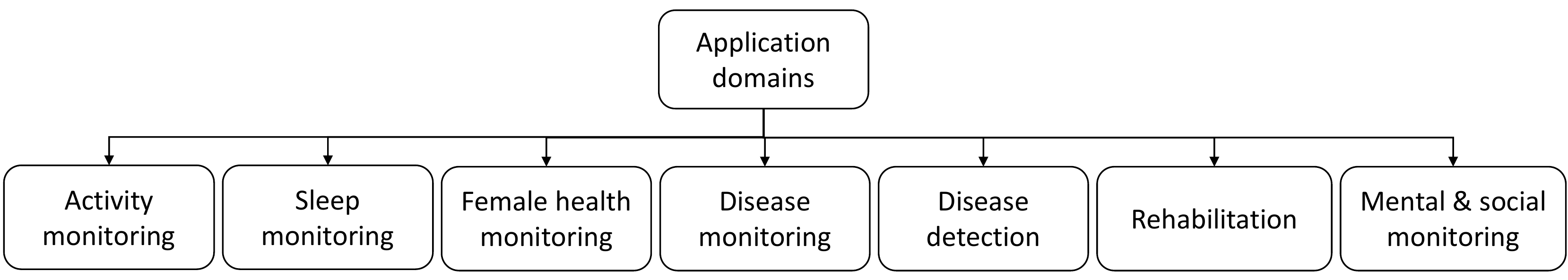}
  \caption{The application domains of in-home health monitoring systems. }
  \Description{}
  \label{fig:app}
\end{figure}

In this section, we review various applications of in-home health monitoring systems. We classify these applications into activity monitoring, sleep monitoring, disease detection, disease monitoring, injury rehabilitation, mental health and social monitoring, and female health monitoring (see Fig.~\ref{fig:app}). In the following sections, we first review each category in detail and then discuss the interconnections between different applications in Subsection \ref{SecSec:inter}.

\subsection{Activity monitoring}
\label{SecSec:activity}
Activity monitoring is one of the most popular research areas in smart home environments \cite{scovanner20073,yu2019pilot,pourpanah2019improved}. It can be used to identify sedentary vs. active lifestyles and encourage active living. It can also be used to detect various incidents or even diseases that might manifest themselves in activity patterns. We categorize the activity monitoring systems into three groups: (\textit{1}) localization of subjects in the home environment, (\textit{2}) daily activity and gait monitoring, and (\textit{3}) fall detection (FD). We will review each of these groups in the following subsections. Table~\ref{table:DA} summarizes the reviewed activity monitoring systems in the context of our study. 

Generally, activity monitoring systems use data obtained from various sensors and process them to identify an individual's physical location or behavior throughout the day~\cite{dodge2015use,bakar2016activity,alam2017besi,mshali2018adaptive,pourpanah2019reinforced,abedi2023ai,rahimi2021self}. Sensors include IMU \cite{pourpanah2021semisupervised,jayalakshmi2020pervasive}, pressure \cite{cantoral2015intelligent,muheidat2020home}, radio frequency \cite{tsukiyama2015home},  WiFi \cite{elgendy2021fog,al2017annotation}, and vision-based \cite{chen2020nonintrusive,mano2016exploiting} sensors. Each of these technologies offers distinct advantages and limitations in monitoring activities within a home environment. IMU sensors provide detailed data on body movements and are suitable for diverse environments. While portability is a significant advantage in such systems, they require user compliance since they need to be worn   \cite{jayalakshmi2020pervasive}. Pressure sensors, such as under-carpet sensors, offer unobtrusive and continuous monitoring without user intervention. These are ideal for stationary monitoring in homes but offer limited movement data and require installation \cite{cantoral2015intelligent,muheidat2020home}. Radio-frequency and WiFi sensors \cite{guo2020healthcare,regani2022wifi} can capture various activities without requiring wearables or environmental alterations. However, their ability to detail specific body movements is generally less precise than IMU systems. Vision-based systems \cite{chen2020nonintrusive,mano2016exploiting} deliver high accuracy in detailed activity monitoring as they rely on cameras and visual sensors \cite{chen2020nonintrusive,mano2016exploiting,luo2018computer}. However, they raise significant privacy concerns as they capture visual data. 
Accordingly, we observe that the choice of technology often depends on the application, type of activity being monitored, and requirements in terms of accuracy, privacy, and user compliance.

\subsubsection{Localization} 
Localization of subjects is crucial for activity monitoring systems as they provide essential information on an individual's movements within the home \cite{awolusi2018wearable}, which can aid in a better understanding of activities. However, indoor localization systems are complex due to the absence of global positioning system (GPS) signals and the lack of direct line of sight in orbiting satellites, which are typically available in outdoor settings~\cite{zafari2019survey}. To overcome this challenge, vision, communication, and pedestrian dead reckoning methods offer an elegant way~\cite{kunhoth2020indoor}. Vision-based systems utilize techniques such as optical flow \cite{zhu2019unsupervised} and 3D reconstruction techniques \cite{kang2020review} to extract information that aids in understanding the location of indoor spaces. Pedestrian dead reckoning-based methods analyze an individual's previous locations using data from accelerometers, gyroscopes, and magnetometers to localize an individual. Specifically, these methods integrate the user's step length, the number of steps taken, and their heading angle to find the position \cite{ban2015indoor}. However, these approaches often accumulate errors over time due to drift and accumulation error \cite{woodman2007introduction}. To mitigate these inaccuracies, many modern navigation systems either combine pedestrian dead reckoning methods with other positioning technologies \cite{zhang2018pedestrian} or employ sensor data fusion techniques \cite{qiu2018inertial} to reduce errors. Communication-based technologies can also be used for indoor positioning, which includes Radio-Frequency Identification (RFID), Wi-Fi, visible light communication, ultra-wideband, and Bluetooth. RFID systems use active or passive tags and determine positions through methods like received signal strength, angle of arrival, time of arrival, and time difference of arrival \cite{lopez2017received,schutz2021wearable}. Wi-Fi systems use existing infrastructure and employ received signal strength fingerprinting or trilateration \cite{he2015wi,regani2022wifi}. Bluetooth systems utilize Bluetooth low energy (BLE) beacons and employ proximity sensing or received signal strength indicator fingerprinting \cite{farid2013recent,bai2020low}. Visible light communication systems use LEDs or fluorescent lights and methods like time of arrival, angle of arrival, and time difference of arrival. Finally, ultra-wideband systems use very short time domain pulses and offer centimeter-level accuracy \cite{do2016depth}.

\subsubsection{Daily activities} 
Monitoring daily activities provides valuable insights in terms of healthcare, safety, and lifestyle management, and enables more informed and proactive approaches to well-being and independence.
The activity and movement recognition system developed in~\cite{bisio2016enabling} utilizes signals from accelerometers and trains ML algorithms for recognition. In~\cite{pais2020evaluation}, DomoCare, which consists of various sensors such as IMUs and ECG, is used to monitor old adults' daily activities such as fridge visits, sleep, and mobility.

Vision-based systems are widely used for activity monitoring \cite{chen2020nonintrusive,mano2016exploiting,luo2018computer}. The study in~\cite{chen2020nonintrusive} integrates a camera and pose estimation models to develop a nonintrusive home monitoring system. It uses feature values of the bounding boxes to characterize the postural changes. 
In~\cite{luo2018computer}, computer vision approaches are leveraged using multimodal visual sensing technologies, such as RGB cameras and thermal sensors, to recognize a wide range of daily activities including sitting, standing, walking, sleeping, using the bedside commode, and getting assistance. Additionally, gait assessment plays a vital role in activity recognition~\cite{roth2018synchronized}. It is the process of evaluating individuals based on the way they walk. Gait assessment involves analyzing various components of an individual's gait, such as walking speed, step width and length, and balance. A review of gait analysis methods using wearable and non-wearable sensors is provided in~\cite{muro2014gait}, while the study in~\cite{sepas2022deep} presents a comprehensive review of gait recognition using DL models.

Pressure sensors installed under carpets or tiles have been used in several studies \cite{cantoral2015intelligent,muheidat2020home,javaid2017balance} to record walking activities. For example, the proposed context-aware real-time in-home monitoring system in \cite{muheidat2020home} installs a sensor pad under the carpet to detect falls, measure gait, and count the number of individuals in the operation area, while in~\cite{cantoral2015intelligent}, plastic optical fiber sensors are installed under the carpet to assess gait and monitor balance.

The study in~\cite{guo2020healthcare} applies signal processing techniques to WiFi channel state information (CSI) for capturing human pose figures and monitoring respiration status. It should be noted, however, that WiFi-based activity recognition methods have limitations in monitoring human activities. On one hand, since the reflection signals from the target are weaker than line-of-sight signals, their range is restricted to a few meters. On the other hand, sensing performance degrades when the target’s reflection signal is combined with interfering reflection signals. 
One possible solution to overcome these issues is beamforming towards the target. To accomplish this, intelligent wireless walls are developed in~\cite{usman2022intelligent} to monitor human activity. This system consists of a reconfigurable intelligent surface for performing beamforming and an ML algorithm, e.g., MLP and RF, for detecting human activities including walking, standing, and sitting.

The study in~\cite{pourpanah2018non} combines Pyroelectric IR sensors with random sampling masks to identify the human body thermal variations and then trains a genetic algorithm-based neural network to recognize human activities, i.e., walking normally or abnormally. 
Multi-person activity monitoring has also been investigated in the literature. For instance, in~\cite{wang2021multi}, a multi-resident (multi-person and pet) activity recognition model for in-home monitoring is proposed. This method first builds a generative model to map sensors in a latent space by maximizing the probability that the subsequent sensor can be read shortly after the current sensor. Then, the data stream is separated into multiple sub-streams, i.e., one for each resident, to recognize activities in multi-resident settings properly. Finally, a single-resident activity recognition model based on a Gaussian mixture probability is used to recognize activities in each sub-stream.
The study in~\cite{nambiar2016low} trains a regression model using data collected from doorway sensors for measuring physical activity levels and tracking movements.
In~\cite{tsukiyama2015home}, water flow sensors are used to measure water usage for three daily activities, including kitchen work, urination, and activities related to maintaining physical cleanliness. Rule-based methods are used to predict health status.

\begin{table*}[t]
\caption{A summary of existing activity monitoring systems. Note that some of the sensors such as the accelerometer, BP, HR, and ECG are embedded in a smartwatch or wristband. In addition, ``--" indicates that the study does not use/mention the corresponding item, i.e., technology or modeling.}
\label{table:DA}
\begin{adjustbox} {width=\columnwidth}
\begin{tabular}{llllll}
\toprule
Study   &Sensor technology & Comm. & Model & Application\\ 
\midrule \midrule
\cite{cantoral2015intelligent} (2015)& Plastic optical fiber & -- & Mathematical model & Gait, Balance monitoring \\ 
\cite{tsukiyama2015home} (2015)& Water flow, RFID tags & Radar & Rule-based model  &  Urination, Physical cleanliness  \\
\cite{ozcan2016autonomous} (2016)& Camera & -- & Threshold & FD\\
\cite{bisio2016enabling} (2016)& Acc. & Cellular & SVM, DTW, DT & Daily activities \\ 
\cite{nambiar2016low} (2016)& Doorway sensors & -- &  Regression &  Daily activities\\ 
\cite{amin2016radar} (2016)& Radar & -- & Mathematical model & FD \\ 
\cite{al2017annotation} (2017)& Wireless & WiFi& Probabilistic model & Daily activities\\ %
\cite{roth2018synchronized} (2018)& Inertial, Pressure &Wireless& Mathematical model & Gait analysis \\
\cite{pham2018delivering} (2018)& Radar, ECG, SpO2, Acc. & ZigBee   & ML & Localization, Daily activities \\
\cite{luo2018computer} (2018)& Camera, Thermal &-- & DL & Daily activities \\
\cite{de2018movement} (2018) & IMU & Wired & Threshold, ML & FD\\
\cite{zhao2018real} (2018)& Depth camera & -- & RF, NN & Detect falls from bed \\
\cite{pourpanah2018non} (2018)& Radar & Wired & MLP  & Motion recognition \\
\cite{bai2020low} (2020)& Acc.  & BLE & Mathematical model &  Localization \\
\cite{guo2020healthcare} (2020)& WiFi signals & WiFi & MLP &  Capture pose figures\\
\cite{chen2020nonintrusive} (2020)& Camera   & Wired & DL & Monitor postural changes \\
\cite{jayalakshmi2020pervasive} (2020) & IMU & Bluetooth & Fuzzy model & Daily activities \\ 
\cite{wu2020fedhome} (2020)&  IMU & WiFi & FL, Generative model &  Daily activities \\ 
\cite{liu2020privacy} (2020)& Camera & --& DL & FD\\ 
\cite{muheidat2020home} (2020)& Floor pressure  & Wired & Cloudlet model &   FD, Count the \# of people \\ 
\cite{elgendy2021fog} (2021)& Camera, ECG, EMG, EEG, Pulse, Temp & Wireless or wired & ML  & Identify patient, Recognize pain \\
\cite{lee2021deep} (2021)& IMU, Camera & -- & DL & FD\\
\cite{vishnu2021human} (2021)& Camera & --& Gaussian model & FD\\
\cite{nimmakayala2021modern} (2021)& HR, BP, PR & Cellular & Mathematical model & Daily activities \\
\cite{liu2021deep} (2021)& Acc. & -- & Deep Autoencoder & FD\\
\cite{wang2021multi} (2021) & IR, Mag., Temp., Light & Wireless & Gaussian model & Daily activities (multi-person) \\ 
\cite{chen2021establishing} (2021) & Camera &-- & FL, R-CNN & Abnormal activities \\ 
\cite{paraschiv2022fall} (2022) & IMU, Depth camera & Wireless & -- & FD\\
\cite{regani2022wifi} (2022)& RFID, WIFi signals & WiFi & Mathematical model & Localization \\ 
\cite{hahm2022home} (2022)&  Acc., Floor pressure  & Wired & Probabilistic models & Gait prediction\\ 
\cite{usman2022intelligent} (2022)& Wireless & Wireless & MLP, RF & Daily activities \\ 
\cite{castillo2022low} (2022)& IMU  & Bluetooth  & ML&  FD \\
\cite{abedi2023ai} (2023) &  Radar & Wireless & GRU, 2D-CNNLSTM  & Daily activities \\ 
\cite{bulcao2023simulation} (2023) & Acc., Camera or Radar& -- & DEVS & FD \\ 
\bottomrule
\end{tabular}
\end{adjustbox}
\end{table*}

\subsubsection{Fall detection} 
\label{SecSec:fall}
Falls are the most prevalent concern among the aging population which can result in severe injuries and even death. The risk of falls not only limits the independence of older adults but also affects their social life and reduces their daily activities. The primary causes that contribute to falls are chronic conditions, weakness, decreased vision and hearing, and side effects of medications. Given the increasing number of fall-related injuries, particularly among the aging population, there is an immediate need to develop in-home FD systems to detect and prevent falls~\cite{mubashir2013survey}. 
Existing FD systems in the literature can be categorized into three main groups: fall detection~\cite{castillo2022low}, fall prediction~\cite{cheng2023incidence}, and fall prevention~\cite{mubashir2013survey}, with the majority of works focusing on FD rather than prediction and prevention.
The study in~\cite{castillo2022low} uses accelerometers and gyroscopes embedded in smartphones/smartwatches to capture signals and transmit the signal by establishing a BLE connection. Then, a mathematical model is designed to process the data and identify falls. The study in~\cite{de2018movement} uses a combination of different signals, such as acceleration, velocity, and displacement, obtained from sensors embedded in wrist-based devices. It applies a set of threshold-based and ML algorithms to identify falls. The study in~\cite{paraschiv2022fall} analyses data acquired from IMUs and depth cameras embedded in a leg band to detect falls in patients with Parkinson’s disease (PD). The FD system proposed in~\cite{liu2021deep} uses a deep Autoencoder structure to analyze low-resolution accelerometer signals.

Examples of vision-based FD systems include the system in~\cite{vishnu2021human} that constructs a Gaussian mixture model to capture motion attributes in both fall and non-fall videos. To achieve this, firstly, a combination of two features, namely the histogram of optical flow and motion boundary histogram features, are extracted, and then, a factor analysis is performed to select low-dimensional and informative representations of the fall motion vector. The proposed FD in~\cite{lee2021deep} trains DL using data collected from RGB cameras and IMUs. 
The study in~\cite{vaiyapuri2021internet} utilizes a CNN-based model to analyze video and detect the occurrence of falls in smart homes.
In contrast to the above-mentioned studies that install cameras at fixed locations, the study in~\cite{ozcan2016autonomous} attaches the camera to the human body. The proposed system employs a histogram of oriented gradients and gradient local binary patterns to extract features, and then a threshold is applied to detect falls. However, the use of vision data raises privacy issues, which may be a point of concern to many adults. To alleviate such concerns, the proposed FD model in~\cite{edgcomb2012privacy} considers five privacy enhancements including applying perturbations such as oval, blur, trailing arrows, box, and silhouette to the videos. Meanwhile, the study in~\cite{liu2020privacy} uses visually shielded video frames to preserve privacy. This method, first, creates a visual shielding effect by applying multi-layer compressed sensing on the image frames. Then, the low-rank sparse decomposition in combination with binary patterns is used to extract object features from the shielded video frames. Additionally, it develops a private information-embedded classification model to accurately identify falls.
Lastly, several studies use depth cameras for FD in the literature~\cite{zhao2018real,bulcao2023simulation}. For instance, the study in~\cite{zhao2018real} uses a depth camera to detect falls from beds. This system, first, detects the human head and upper body using RF, and then the nearest neighbor classifier is utilized to detect falls from the bed using the motion information of the human upper body.   
In~\cite{bulcao2023simulation}, a simulation model based on discrete event system specifications (DEVS) for FD is presented. This study compares the performance of several architectures that are formed by combining various wearable sensors, e.g., accelerometers embedded in smartphones or smartwatches, with depth cameras to detect falls.

\subsection{Sleep monitoring}
\label{SecSec:sleep}
High-quality sleep is crucial for health and well-being, and sleep disorders have been highly correlated to various health outcomes~\cite{dregan2011cross}. 
The International Classification of Sleep Disorders classifies sleep disorders into seven categories. These categories include insomnia, hypersomnolence, parasomnia, circadian rhythm sleep-wake, sleep-related breathing and movement, and other disorders~\cite{sateia2014international}. 
In the long term, sleep disorders can have health implications and increase the probability of developing various diseases such as cardiovascular, metabolic, and psychiatric disorders. Moreover, poor-quality sleep can deteriorate mental and physical performance~\cite{friedrich2022assistive,pais2020evaluation}. Therefore, the development of sleep monitoring systems to continuously monitor and assess the individual's health is a crucial area in the context of in-home health monitoring. A summary of existing sleep monitoring systems is provided in Table~\ref{table:sleep}.

\begin{table*}[t]
\caption{A summary of existing sleep monitoring systems. Note that some of the sensors such as the accelerometer, BP, HR, and ECG are embedded in a smartwatch or wristband. In addition, ``--" indicates that the study does not use/mention the corresponding item, i.e., technology or modeling.}
\label{table:sleep}
\begin{adjustbox} {width=\columnwidth}
\begin{tabular}{llllll}
\toprule
Study (year)   &Sensor technology & Comm. & Modeling & Application\\ 
\midrule \midrule
\cite{gong2015home} (2015) & Acc., Audio  & Wireless & --& Detecting incontinence, sleep agitation\\
\cite{lin2016sleepsense} (2016)& Acc., Air flow, Radar  & Wireless & DT & Sleep status\\ 
\cite{hsu2017zero} (2017)& RF, Audio &Wireless & Hidden Markov Mode, DNN & Monitoring insomnia, sleep \\ 
\cite{sadek2018nonintrusive} (2018)& Fiber optic & Wired & -- & Sleep motoring \\
\cite{moon2019integrated} (2019) & ECG, IMU & BLE & -- & Monitoring nocturnal enuresis during sleep \\
\cite{pais2020evaluation} (2020)&  PIR, ECG & Wireless & Mathematical model & Sleep, Daily activities\\
\cite{kristiansen2021machine} (2021)& PSG & Wired & Shallow and deep models  & Sleep Apnea detection \\ 
\cite{pan2021home} (2021) & Acc. & -- & Threshold, $K$-means & Sleep monitoring \\
\cite{honda2022wearable} (2022)& Humidity sensors & Bluetooth & -- & Diagnose sleep apnea symptoms \\
\bottomrule
\end{tabular}
\end{adjustbox}
\end{table*}

Polysomnography (PSG) is considered a gold standard technique in sleep research. This technique monitors the patient's sleep with EMG, ECG, EEG, pulse, and respiration to measure sleep quality, structure, and disturbance. In contrast, the polygraphy (PG) technique is a simplified version of PSG that does not measure/monitor the sleep structure. Although PSG is a comprehensive technique for sleep monitoring, it is resource-intensive, expensive, and usually requires patients to be administrated in hospitals or sleep centers~\cite{vandenberghe2015sleep,kristiansen2021machine}. In addition, both PSG and PG require various signal measurements. Therefore, these techniques are not feasible to administer in in-home settings.
To overcome these issues, researchers have attempted to develop simple and non-intrusive systems that are easy to use by individuals and suitable for in-home environments. For example, the study in~\cite{pan2021home} monitors sleep by measuring wrist movement using a 3-axis accelerometer. The proposed sleep monitoring system in~\cite{sadek2018nonintrusive} uses an optical fiber embedded sensor mat under the mattress to continuously monitor the quality of sleep based on sleep duration and vital signs. Optical fiber mats are highly sensitive to physical movements, breathing, and HR. Their use ensures a high level of privacy as no visual or sound data is captured, but their monitoring is confined to the bed area, making them ineffective if the user is not on the mattress. Pressure sensors embedded in beds also provide a privacy-preserving approach for pose estimation during sleep.  
They offer a non-intrusive method to analyze body positions and movements during sleep. These methods use custom-designed \cite{davoodnia2019identity,liu2023pressure} or off-the-shelf pose estimators \cite{davoodnia2021bed,davoodnia2022estimating} to detect the general pose (e.g., supine, left, or right) \cite{davoodnia2019identity} or fine-grained 2D/3D pose information \cite{davoodnia2021bed, davoodnia2022estimating,liu2023pressure}.

SleepSense~\cite{lin2016sleepsense} uses a radar sensor to continuously obtain sleep-related signals from the patient and recognize the sleep status. Radar sensor-based systems allow for non-contact sleep monitoring, detecting movements and vital signs without the need for direct physical contact or specific positioning. This approach offers flexibility in monitoring, regardless of the user's location in the room, while also maintaining privacy. However, these systems may not match the precision of direct contact methods in distinguishing specific sleep stages or detailed movements and can be influenced by other environmental factors. 
EZ-Sleep~\cite{hsu2017zero}, which is a contactless sleep sensor for monitoring insomnia and sleep, utilizes RF signals that reflect off the user's body to record parameters such as total sleep time, sleep latency, and sleep efficiency. EZ-Sleep can also automatically identify the location of users and simultaneously monitor multiple users. In~\cite{kwon2021recent}, the respiration rate is used to monitor sleep apnea syndrome. Notably, various sensors, such as pressure, humidity, accelerometers, and strain sensors, can be employed to measure respiration rates~\cite{honda2022wearable}. 
Finally, the study in~\cite{moon2019integrated} develops a sensor-based framework consisting of electrical impedance tomography, ECG, and IMU to monitor body impedance values, HR, and periodic limb movements during sleep, for health applications such as bedwetting.

\subsection{Disease detection}
This category of monitoring systems aims to provide continuous surveillance of individuals and identify various diseases as early as possible and provide timely interventions before they become major problems. Table~\ref{table:DD} reviews disease detection systems. Recently, the development of such systems has become important due to factors such as the aging population, epidemics, and pandemics such as COVID-19.  Fever is one of the common issues that can be caused by a variety of underlying conditions such as infections, inflammatory disorders, medication, paraplegia, and others~\cite{rabadi2021fever}. In particular, with pandemics such as the COVID-19 pandemic, fever has become an important symptom to detect as it can be an indicator of the virus. The study in~\cite{al2021design} develops a WSN to continuously monitor BT and alert healthcare providers when a fever is detected. The study in~\cite{rahman2021internet} develops a DL model to detect COVID-19 symptoms.

Several studies attempt to continuously monitor vital signs such as respiration, HR, BP, BT, SpO2, and etc~\cite{adib2015smart,zhang2020accurate,dharmik2021iot,siam2022portable,xie2022passive,feng2017non,teja2018smart,anzanpour2015internet,bertsch2020design,am2021survey,alazzam2021novel,guo2021emergency}, which can all be used to detect various diseases in different stages of progression. For example, the study in~\cite{feng2017non} uses ultrasensitive accelerometers to record ballistocardiogram data for measuring HR and respiration, while the proposed system in~\cite{teja2018smart} uses ECG signals and a pulse oximeter to measure HR and monitor vital signs. 
In~\cite{guo2021emergency}, WiFi signals are employed to continuously monitor the position, behavior, and respiration of older adults. It uses the WiFi CSI to extract the emergency semantic feature vectors. First, the WiFi signals are converted into two keypoint maps and respiration graphs, which are then used to extract features related to the position/behavior and respiration, respectively. In addition, Vital-Radio~\cite{adib2015smart} uses wireless sensing technology embedded in smart homes to track breathing and measure HR. It computes the changes in the distance between the human body and the wireless sensor, as the distance slightly changes during inhale and exhale. The distance is measured by computing the time that a low-power wireless signal travels to the human body and returns to its antenna. It then uses the Fourier transform to extract the frequency rate for monitoring vital signs. Vital Radio can simultaneously track the vital signs of multiple people. 
In contrast, the study in~\cite{xie2022passive} uses UWB and depth sensors. After identifying four vital temporal and spectral signs as well as their vital sign estimator, a probabilistic weighted framework is used for tracking vital signs. Additionally, a two-branch LSTM-based model is developed to monitor multiple individuals and distinguish between their activities at home.

Furthermore, the method in~\cite{zhang2020accurate} uses RFID technology to perform continuous respiratory monitoring of moving people. However, there are two challenging issues in developing such a monitoring system. The first challenge is that small movements associated with breathing can be obscured by body movements. The second challenge is the phase ambiguity that occurs when the body is moving, which is more complex than when the body is static. To address the first challenge, a pair of tags are used to mitigate the effects of the individual’s body movement. While a distance-tracking model is developed for monitoring the phase transitions during user movement to handle the phase ambiguity issue.
The study in~\cite{bhowmik2022iot} proposed to monitor both environmental conditions and vital signs. For monitoring the environmental conditions, it measures the carbon monoxide level at home. It also attempts to aggregate data from various sources and handle the missing data. 
The study in~\cite{hossain2016cloud} employs a one-class SVM to identify abnormal ECG signals, which could be indicators of certain diseases. The study in~\cite{wu2021early} identifies abnormal conditions by analyzing the trajectory of data streams including motion, depth, and bed sensors. It then extracts features to train a sequential probabilistic Gaussian mixture model~\cite{wu2019data} to track the health trajectory and predict abnormal conditions in its early stages. 

It should be noted that wearable systems can also be widely used for the detection of abnormal conditions in vital signs, which could be indicative of certain diseases. However, a detailed review of wearables is beyond the scope of this study. For a review on wearable technologies, please see~\cite{loncar2019literature,ferreira2021wearable}.

\begin{table*}[t]
\caption{A summary of disease detection systems. Note that some of the sensors such as the accelerometer, BP, HR, and ECG are embedded in a smartwatch or wristband. In addition, ``--'' indicates that the study does not use/mention the corresponding item, i.e., technology or modeling.}
\label{table:DD}
\begin{adjustbox} {width=\columnwidth}
\begin{tabular}{lllllll}
\toprule
Study (year)  &Sensor technology & Comm. & Model & Application \\
\midrule \midrule
\cite{adib2015smart} (2015)& Radar & Wireless & Mathematical model & tracking breath, HR \\
\cite{hossain2016cloud} (2016)& ECG & Cellular & One class SVM & Identify abnormal ECG \\
\cite{feng2017non} (2017)& Ultrasensitive Acc. & -- & Autocorrelation & HR and respiration \\
\cite{verma2018fog}(2018)& Various sensors & -- &Bayesian belief network  & Situation of the patient \\ 
\cite{conn2019home} (2019) & ECG &Wireless& --  &  Cardiovascular monitoring \\
\cite{allen2019sms} (2019)& Audio & SM &-- & BP\\
\cite{zhang2020accurate} (2020)& RFID & Wireless & Mathematical model & Respiration monitoring \\
\cite{bertsch2020design} (2020)& Acc., PPG& -- & - & Monitoring vital signs \\ 
\cite{guo2021emergency} (2021)& WiFi signals& WiFi & Mathematical model & Position, Behavior, Respiration\\ 
\cite{dharmik2021iot} (2021)& Temp., Pulse & Wired, Wireless & DT, KNN, K-means & Vital signs \\
\cite{wu2021early} (2021) & IR, Depth camera, Hydraulic bed &-- & Gaussian model & Tracking health trajectory \\
\cite{flaucher2022smartphone} (2022) &  Camera   & --&CNN  & Prenatal care \\
\cite{siam2022portable} (2022) & PPG, SpO2, ECG, Temp., Humidity & Wireless & Mathematical model & Vital signs \\
\cite{bhowmik2022iot} (2022)& BT, Humidity, SpO2, CO & Bluetooth & Mathematical model & Monitoring health, Environment \\
\cite{xie2022passive} (2022) & Depth sensors  & UWB & LSTM & HR, Respiratory rates  \\
\cite{farooq2022blockchain} (2022)& IMU, ECG  & Wireless & FL & Health monitoring \\ 
\bottomrule
\end{tabular}
\end{adjustbox}
\end{table*}

\subsection{Disease monitoring} 
Whereas disease detection systems aim to identify sudden changes in healthy individuals that could indicate a new disease, in-home disease monitoring systems aim to monitor the progress of different `known' diseases and the effectiveness of medical interventions. Additionally, they can produce emergency alarms or notify healthcare professionals in critical situations. Table~\ref{table:DM} summarizes the literature on in-home disease monitoring systems. 

\begin{table*}[t]
\caption{A summary of existing disease monitoring systems. Note that some of the sensors such as the accelerometer, BP, HR, and ECG are embedded in a smartwatch or wristband. In addition, ``--" indicates that the study does not use/mention the corresponding item, i.e., technology or modeling.  }
\label{table:DM}
\begin{adjustbox} {width=\columnwidth}
\begin{tabular}{lllll}
\toprule
Study (year)  &Sensor technology & Comm. & Model & Application\\ 
\midrule\midrule
\cite{gong2015home} (2015) & Acc., Audio  & Wireless & --& Detecting incontinence\\ 
\cite{boletsis2015use} (2015) & BT, Acc., Blood flow  & WiFi & -- & Monitoring dementia\\
\cite{medina2017real} (2017) & ECG, BP, HR   & -- & Fuzzy model  & Cardiac rehabilitation \\
\cite{miao2018wearable} (2018) & ECG, PPG  &Bluetooth & MLR, DT, BP & Arterial stiffness evaluation\\
\cite{enshaeifar2018health} (2018) & BP, HR, Temp., Environmental   & -- &  Markov chain & Monitoring dementia \\
\cite{chatterjee2018designing} (2018) & Multiple sensors &  Wireless & ANN & Chronic disease of diabetes\\

\cite{lv2020teleoperation} (2020) & Camera & Ethernet & Mathematical model & Dementia \\
\cite{enshaeifar2020digital} (2020) & RFID & Bluetooth & ML algorithms & Identifying the risks of health \\
\cite{naeem2020cnn} (2020) & Camera & --& CNN & Pill reminder  \\
\cite{martiradonna2021cascaded} (2021) & EEG, camera, ECG, EMG, SpO2& Cellular &-- &  Epileptic patients\\ 
\cite{keikhosrokiani2022iot} (2022) &BP, HR, ECG, PPG, SpO2  &Bluetooth, WiFi  & DT  & Remote heart monitoring \\
\cite{fu2022short} (2022) & BT, BP, SpO2 & Bluetooth & -- & After lung cancer surgery \\
\bottomrule
\end{tabular}
\end{adjustbox}
\end{table*}

Neurological conditions such as dementia, Alzheimer’s Disease (AD), PD, and epilepsy have been widely targeted for in-home monitoring in the literature 
~\cite{moshnyaga2015identification,enshaeifar2018health,enshaeifar2020digital,boletsis2015use,martiradonna2021cascaded,gong2015home}. 
For instance, the study presented in~\cite{enshaeifar2018health} extracts patterns that are directly correlated with irritability, agitation, and aggression, in patients with dementia. It uses ML techniques to explore the relationship between environmental and physiological data for monitoring patients and discovering any changes in their health. The work presented in~\cite{gong2015home} analyzes the relationship between urinary incontinence and sleep continuity and the number of occurrences of nighttime agitation for patients with AD. This study employs various sensors to detect wetness, nighttime agitation, and speech outbursts. Another study~\cite{enshaeifar2020digital} uses ML and analytical algorithms to identify the risks of adverse health conditions such as urinary tract infections. Meanwhile, the study in~\cite{moshnyaga2015identification} monitors and localizes people with dementia using various sensors and ML models. The proposed framework in~\cite{martiradonna2021cascaded} continuously monitors epileptic patients in their homes by performing two services: (\textit{1}) ongoing monitoring, and (\textit{2}) emergency response which can be activated during life-threatening situations.

Cardiovascular monitoring is another essential aspect of in-home health monitoring, particularly for individuals who are at risk of developing cardiovascular disease or those who have already been diagnosed with the condition~\cite{conn2019home}.
To this end, several methods have been developed, each offering unique benefits and facing distinct challenges for in-home health monitoring. The wrist-worn device described in~\cite{medina2017real} integrates an HR sensor with a cardiac rehabilitation program, offering real-time monitoring of physical activities. Its use of a fuzzy model to simulate the cardiac rehabilitation protocol provides personalized care but may require user engagement in consistently wearing the device. RemoteHeart~\cite{keikhosrokiani2022iot} utilizes ML models to analyze data from a range of biomedical sensors, including blood pressure, SpO2, ECG, PPG, and HR. This system offers a comprehensive view of heart health remotely, making it highly beneficial for continuous monitoring, though it may face challenges in data management and interpretation. The study in~\cite{okeke2019technology}  discusses the impact of technology on home health aids in the context of heart failure management.
The developed in-home arterial stiffness evaluation device in~\cite{miao2018wearable} extracts various features from ECG and PPG for training various ML models such as multivariate linear regression, DT, and BP neural networks. Wearable ECG sensors provide a direct and user-friendly approach to cardiovascular monitoring, suitable for continuous observation. They are less intrusive than traditional medical equipment, but their effectiveness depends on the precision of sensors and algorithms used. Reviews in~\cite{shabaan2020survey,nigusse2021wearable} provide further insights into these wearable systems, underscoring their potential in the early detection and long-term management of cardiovascular diseases. 
Given the challenges of recording ECG through wearable devices, studies such as \cite{sarkar2021cardiogan,shome2024region} have proposed the translation of PPG to ECG, which allows existing ECG frameworks to be used for cardiovascular health monitoring based on PPG signals. These studies have demonstrated benefits in utilizing the synthesized ECG as opposed to the original PPG signals in terms of cardiac disease detection.

Several studies focus on detecting health changes in individuals with chronic conditions, each employing different technologies and techniques~\cite{skubic2015automated,chatterjee2018designing,allen2019sms,chatrati2022smart,persell2018design}. The study in~\cite{skubic2015automated} proposes using IR sensors to capture motion signals and trains multiple one-class classification models to recognize abnormal patterns. 
The studies in~\cite{chatterjee2018designing,chatrati2022smart} focus on diabetes management by training supervised learning models to predict blood glucose levels. These studies demonstrate the potential of ML in managing diabetes and offer personalized insights into blood sugar trends. 
The works in~\cite{persell2018design,allen2019sms} develop phone- and SMS-based systems for monitoring hypertension and blood pressure. These systems are particularly user-friendly and accessible, making them suitable for a wide range of users, including those not technologically adept. However, their effectiveness is contingent upon user engagement and accurate self-reporting.
Additionally, the work in~\cite{rahman2016gesture} introduces a gesture-based model to assist individuals with physical impairments. It showcases the use of technology in improving the quality of life for people with mobility challenges. While innovative, the practicality and user acceptance of such gesture-based systems can vary greatly among individuals.

Furthermore, the study in~\cite{fu2022short} focuses on monitoring lung diseases. In this work, a remote monitoring system is developed for post-lung cancer surgery patients. This system integrates Telehealth electronic monitoring with a smartphone application, guiding patients and tracking vital parameters like BT, BP, SpO2, and others. The data is uploaded to a platform for medical staff review, offering real-time monitoring and potentially improving postoperative care. The study in~\cite{wijsenbeek2023home} reviews in-home monitoring systems for interstitial lung diseases, providing insights into current practices and future directions. Their review covers advanced procedures as well as off-the-shelf systems such as pulse oximeters, cough monitors, and activity trackers, which offer comprehensive and continuous monitoring options that may improve disease management. However, the challenge lies in aggregating these various data sources for a holistic view of the patient's condition, and ensuring that these technologies are accessible and user-friendly for patients at home.

\subsection{Rehabilitation}
These systems have been developed to provide real-time monitoring and tracking of patients' progress during rehabilitation. In-home injury rehabilitation monitoring systems can provide real-time feedback to patients recovering from injuries and physical impairments and ensure that they correctly perform their therapy. In addition, these systems allow in-home therapy to patients and receive instructions from their therapists. A summary of injury rehabilitation systems is presented in Table~\ref{table:Rehab}.  
The proposed in-home therapy monitoring system in~\cite{rahman2017mtherapy} uses multiple gesture-tracking sensors to identify various joints as well as their movements in both rotational and angular directions. The study in~\cite{chiang2018kinect} develops a Kinect-based intervention system to improve individuals' exercise at home. This model, first, uses a Kinect sensor and a motion capture system to simultaneously record the joint position, and then, maps Kinect measurements to the motion capture measurements by training a Gaussian process model.

Studies~\cite{park2017development,gomez2022monitoring,sanders2020feasibility,lee2018enabling,fang2020longitudinal,forbrigger2023considerations} focus on developing rehabilitation devices to monitor the movement of various body parts post-stroke.
The study in~\cite{gomez2022monitoring} develops a device named ARM to monitor arm movement. It employs classification algorithms to analyze signals from an IMU worn on the wrist. ARM collects two types of data: functional assessment data involving standard movements and data from activities of daily living, like pizza-making. This approach allows for a comprehensive understanding of the patient's abilities in both controlled and real-life scenarios. The study in~\cite{sanders2020feasibility} introduces MusicGlove, a sensor-equipped glove that uses music-based therapy for in-home finger rehabilitation. 
It combines therapeutic exercises to enhance patient motivation and adherence to rehabilitation protocols in the early stages following a stroke. 
The works in~\cite{lee2018enabling,fang2020longitudinal} focus on wearable devices for limb recovery. The developed device in~\cite{lee2018enabling} uses sensors to monitor stroke-affected upper limbs, providing feedback to encourage the use of the affected limb. In contrast, the telerehabilitation system in~\cite{fang2020longitudinal} offers a remote but personalized rehabilitation approach by measuring limb mobility and classifying patient impairment levels. 

\begin{table*}[t]
\caption{A summary of the existing injury rehabilitation systems. Note that some of the sensors such as the accelerometer, BP, HR, and ECG are embedded in a smartwatch or wristband. In addition, ``--" indicates that the study does not use/mention the corresponding item, i.e., technology or modeling.  }
\label{table:Rehab}
\begin{adjustbox} {width=\columnwidth}
\begin{tabular}{lllll}
\toprule
Study (year)  &Sensor technology & Comm. & Model & Application\\ 
\midrule \midrule
\cite{rahman2017mtherapy} (2017)& Gesture-tracking sensors &  Wi-Fi, ZigBee & -- & Therapy  \\
\cite{chiang2018kinect} (2018)& Kinect sensor, Motion capture & Wireless & Regression  &Improve individuals’ exercise \\
\cite{lee2018enabling} (2018)& Acc. & Bluetooth & Regression & Upper limb Rehabilitation\\
\cite{sanders2020feasibility} (2020)& Grip sensor & --& -- & Finger rehabilitation after stroke \\ 
\cite{fang2020longitudinal} (2021)&  Acc. & ZigBee & Fuzzy model & Limb function recovery after stroke\\
\cite{gomez2022monitoring} (2022)& Acc.  & Bluetooth &CNN, RF & Arm rehabilitation after stroke\\
\cite{lee2022performing} (2022)&  Visual, Vibrotactile & Cellular or WiFi & Mathematical model&  Balance, stability\\ 
\bottomrule
\end{tabular}
\end{adjustbox}
\end{table*}

\subsection{Female health monitoring}
\label{SecSec:female}
In-home health monitoring can provide convenient and accessible healthcare services for females. These services are particularly important for those who may face barriers to accessing conventional healthcare facilities. These systems enable women to receive regular checkups and monitor their health from home, which can lead to better health outcomes and a more proactive approach to managing their health. Specifically, these systems can be beneficial for women during pregnancy~\cite{kazantsev2012development} and postpartum care~\cite{thomas2021patient}, menstrual cycle tracking~\cite{maijala2019nocturnal}, and managing other conditions.
The study in~\cite{kazantsev2012development} highlights the use of in-home monitoring during pregnancy. It offers a way for expectant mothers to track their health and the baby's development without frequent hospital visits. The study in~\cite{thomas2021patient} focuses on postpartum care, emphasizing the importance of accessible healthcare support during this critical period. The work in~\cite{maijala2019nocturnal} explores menstrual cycle tracking, providing women with valuable insights into their reproductive health. Furthermore, \cite{flaucher2022smartphone} proposes a vision-based approach for analyzing urine at home during prenatal care to identify pregnancy-related diseases. To achieve this, various images are obtained from an in-home control urine study. They are then labeled and used to train a region-based CNN to perform the prediction of pregnancy-related diseases. This approach offers a non-invasive, user-friendly method for early disease detection, although it requires accurate image capture and processing. Overall, these in-home health monitoring systems present the potential for advanced in-home women's health monitoring systems by offering personalized, convenient, and non-invasive options for health management. They represent a shift towards more patient-centric healthcare that could empower women to actively participate in their health tracking and decision-making. 

\begin{table*}[t]
\caption{A summary of existing mental health and social interaction monitoring systems. Note that some of the sensors such as the accelerometer, BP, HR, and ECG are embedded in a smartwatch or wristband. In addition, ``--" indicates that the study does not use/mention the corresponding item, i.e., technology or modeling.}
\label{table:mhm}
\begin{adjustbox} {width=\columnwidth}
\begin{tabular}{llllll}
\toprule
Study (year)  &Sensor technology & Comm. & Model & Application \\
\midrule \midrule
\cite{lyons2015pervasive} (2015)& IR, Wireless & Zigbee & Mathematical model & Mood, Cognitive function \\
\cite{dickerson2015empath2} (2015)& Scales, PIR, Acc., Audio & Bluetooth, Zwave& Inference trees & Depression, sleep/stress, Agitation  \\
\cite{miura2020implementing,miura2020empirical,miura2019prototyping} (2020)& Mind sensing~\cite{nakamura2019developing} & Cellular & Mathematical model & Monitoring mental status \\
\cite{gao2020monitoring} (2020)& Audio& -- & Recommendation System & Caregiver's mood state\\
\cite{bianco2021smart} (2021)& Camera, Audio& -- & DL & Emotion recognition  \\
\cite{alazzam2021novel} (2021)& PPG, BP &  Wireless & MLP, DT & BP, stress monitoring\\ 
\cite{behinaein2021transformer} (2021)& ECG  & -- & Transformer & Stress detection \\
\bottomrule
\end{tabular}
\end{adjustbox}
\end{table*}

\subsection{Mental health and social interaction monitoring}
\label{SecSec:mental}
The mental well-being of individuals is essential in promoting sustainable in-home long-term care and encouraging their ability for self-care. In this regard, several mental health monitoring systems have been developed. 
The work in \cite{behinaein2021transformer} develops a transformer-based neural network capable of accurate detection of stress from ECG, offering the potential to be used in wearable or mobile devices. The works in~\cite{miura2020empirical,miura2019prototyping} propose mind monitoring services using a chatbot that assesses mental health by analyzing responses to user-specific questions. This approach leverages interactive technology to engage users and track their mental state over time, though it relies heavily on user participation and accurate self-reporting. The study conducted in~\cite{gao2020monitoring} develops a framework aimed at enhancing interactions between dementia patients and caregivers in home environments. This system includes a mood monitoring component based on speech analysis and a recommendation system to alleviate caregiver stress. 
The study in~\cite{lyons2015pervasive} monitors behaviors such as gait, sleep patterns, and medication adherence by installing sensors in over 480 homes for more than eight years. This extensive study provides insights into factors like cognitive functions and mood through a significant investment in long-term data collection and analysis.

Empath2~\cite{dickerson2015empath2} is an in-home healthcare architecture designed to monitor depression and the relationship between sleep, stress, and nighttime agitation. Its adaptation to diverse mental health conditions demonstrates versatility, though the complexity of monitoring multiple factors simultaneously can be challenging. The developed smart mirror in~\cite{bianco2021smart} detects and analyzes motion using visual and audio interactions. It employs DL models for face detection, identification, and emotion recognition. This approach offers an integrated solution for monitoring mental health but raises privacy concerns due to its visual and audio data processing.
Table~\ref{table:mhm} provides a summary of mental health and social interaction monitoring systems.

\subsection{Interconnections between applications}
\label{SecSec:inter}
There exist various interesting and intricate interconnections between different applications of in-home health monitoring systems. For example, the data collected from activity monitoring systems could also be essential for injury rehabilitation, disease detection, disease monitoring, and mental health monitoring. In the context of rehabilitation, activity data can help healthcare providers assess patients' progress, track movement patterns, and determine the effectiveness of rehabilitation plans \cite{ferraris2014remote}. For disease detection and monitoring, changes in activity levels and patterns can serve as early indicators of potential health issues or disease progression \cite{hallal2012global}. Moreover, in the context of mental health, changes in activity levels and patterns can offer insights into an individual's mental well-being, with changes often correlating with fluctuations in mood, energy levels, and overall psychological state \cite{jiang2018wearable}. Conversely, data from both sleep and mental state of individuals can be used to contextualize information for various applications such as disease monitoring \cite{rowan2022future,yvellez2018p051}.

\section{Future Research Directions}
\label{Sec:discussion}
In this survey, we have been delving into various types of ubiquitous in-home health monitoring systems and identifying their main components, including advanced sensing technologies, communication technologies, and intelligence and computing systems. 
Our findings indicate that despite the considerable progress in developing in-home health monitoring systems, many of them have only shown success in controlled laboratory environments and have not yet produced reliable results when implemented in real-world settings. Therefore, we strongly argue that there is still a substantial amount of research needed to build effective and reliable in-home health monitoring systems. To address this challenge, it is important to identify research gaps and avenues for further investigation. These gaps highlight areas where existing research has not yet fully addressed the challenges of in-home health monitoring systems. In the following sections, we highlight the most important research gaps and provide potential avenues for further research.

\subsection{Robust and easy-to-use sensor design}
Sensors play a key role in developing in-home health monitoring systems as they are used to collect a wide range of health and environmental data.
They must have several characteristics, such as accuracy, reliability, sensitivity, low power consumption, durability, and cost-effectiveness, to establish accurate data acquisition \cite{hasan2020wearable}. 
Contact sensors embedded in wearable devices are more accessible, cost-effective, and can obtain more accurate data~\cite{chiang2018kinect}. However, a common trade-off exists between the wearability and ease of use of these sensors, and the quality of data they produce. This is particularly evident when comparing them to lab-grade equipment, which, despite its bulkiness and complex setup requirements like calibration procedures, often provides higher quality data \cite{uddin2020body}. Another significant challenge with sensors is their susceptibility to noise and variations in real-world conditions \cite{heikenfeld2018wearable}. These factors can lead to distorted or inaccurate sensor measurements, thereby compromising the quality and reliability of the data collected. Additionally, ensuring a constant power supply poses a substantial challenge, as sensors and data processing units in these systems often have high power requirements \cite{zhu2015wearable}. Lastly, the lack of standardization in assessing the quality of different sensors presents a critical challenge. This absence of a unified standard makes it difficult to analyze and compare the performance of various in-home health monitoring systems \cite{pantelopoulos2009survey}. Addressing these challenges is essential for the advancement and reliability of in-home health monitoring technologies.

\begin{tcolorbox}[colback=white!98!blue,colframe=blue!40!black]
\textbf{Recommendation 1:} We believe that there are several research directions in the field of sensors for robust in-home health monitoring systems that require further investigation. The primary objective in this context is to improve the usability of sensing devices without compromising quality or durability. Accordingly, developing new sensors that offer lab-grade signal quality in wearable and user-friendly form factors, with minimal power consumption, is essential. Second, it is essential to develop standardized measures and techniques to quantify and assess the quality and consistency of sensor data, which could be used to determine the reliability of the analyses performed by different health monitoring systems. Lastly, creating universal standards for different types of sensors can be an essential step toward enabling the aggregation of data across platforms and products.
\end{tcolorbox}

\subsection{Responsible artificial intelligence}
Responsible artificial intelligence (RAI) refers to the design and development of AI systems that are ethical, transparent, robust, and fair while adhering to strict privacy and safety guidelines \cite{shahriari2017ieee,muller2020ethics,arrieta2020explainable,dignum2019responsible}.
Explainability and interpretability are crucial for in-home health monitoring systems, especially when using complex ML and DL algorithms \cite{schlegel2019towards}. They enable healthcare providers and patients to understand the rationale behind system predictions or decisions. 
Additionally, addressing biases and promoting fairness in ML and DL models is vital due to potential inherent biases in health data, which may adversely affect the accuracy and fairness of the resulting models \cite{dwork2011fairness}. 
This can arise in various situations, such as in imbalanced datasets, or towards certain sensitive attributes, including gender, age, ethnicity, and race~\cite{dwork2011fairness}. 
Robustness is another key factor in developing in-home health monitoring systems. These systems must be able to perform without failure under a wide range of conditions, such as various distribution shifts \cite{soltanieh2023distribution}, noise \cite{cooper2022accountability}, and uncertainty \cite{zhou2022survey,abdar2021review}, which require robustness in their design and implementation.
Lastly, the privacy of health data is critical due to its sensitivity and potential consequences like identity theft and confidentiality breaches \cite{hadian2016privacy}. 
As a result, given the sensitivity of health data and AI systems, addressing these challenges is essential for the reliability of in-home health monitoring systems. Therefore, we recommend that the research community focus on integrating RAI into in-home health monitoring systems.

\begin{tcolorbox}[colback=white!98!blue,colframe=blue!40!black]
\textbf{Recommendation 2:} Incorporating RAI into the development of in-home health monitoring systems is a crucial consideration in automated or AI-assisted healthcare. We strongly recommend that future research in this field focus on addressing these concerns, i.e., security and privacy, explainability and interoperability, bias and fairness, and trustworthiness, to ensure the safe and effective implementation of RAI-powered health monitoring systems.
\end{tcolorbox}

\subsection{Real-time analysis} 
Most existing in-home health monitoring systems rely on offline analysis, which involves collecting data from sensors and analyzing it post-collection. 
However, offline analysis is inherently limited in providing real-time feedback and intervention.  
Several studies such as \cite{amirshahi2019ecg,majumder2019real,raj2020efficient} highlight these limitations, particularly in cardiac monitoring systems where delays in essential interventions can occur. Similarly, research in \cite{el2018mobile,kumar2021advanced} points to significant delays in insulin adjustment in diabetes management systems due to offline data analysis. 
Moreover, research such as \cite{larkai2015wireless,reddy2021smart} in elderly care and FD systems underlines the importance of real-time data processing. These studies demonstrate how immediate data analysis can significantly mitigate injury risks by enabling prompt responses.
Given these challenges, it is imperative to develop real-time health monitoring systems that can analyze data and provide timely feedback and intervention.

\begin{tcolorbox}[colback=white!98!blue,colframe=blue!40!black]
\textbf{Recommendation 3:} The integration of edge computing and fog computing, 5G+ communications, and the implementation of online or continual learning models, can significantly improve the real-time analysis and feedback of in-home health monitoring systems. Such improvements have the potential to greatly enhance patient outcomes and their quality of care.
\end{tcolorbox}

\subsection{Prediction and forecasting instead of detection}
We notice that most of the existing in-home health monitoring systems in the literature perform detection rather than prediction/forecasting \cite{castillo2022low,paraschiv2022fall,vaiyapuri2021internet,vishnu2021human,lee2021deep}. However, developing `predictive' in-home health monitoring systems is a key missing piece as it can provide significant benefits to both patients and healthcare providers \cite{ponnapalli2023triple,sanchez2016use}. To elaborate on this notion, detection includes monitoring individuals' health data and identifying abnormalities in the event of occurrence. For example, an FD model continuously monitors individuals and warns if it detects a fall. In contrast, prediction/forecasting includes employing data analysis methods to examine trends in data and predict the probability of a specific disease occurring in the future. By predicting adverse health events, proactive steps can be taken to prevent them \cite{chatrati2022smart}. 
In fact, recent advancements underscore the potential of predictive systems in enhancing patient care through timely interventions and reducing hospital admissions \cite{bedoya2023explainability,bhowmik2022iot}. Furthermore, the integration of cloud, edge, and fog computing in health monitoring systems \cite{ponnapalli2023triple} as well as time-series forecasting DL algorithms \cite{yue2022ts2vec,lim2021time}, highlights the evolving technological landscape that can support more sophisticated predictive analytics in healthcare.
Therefore, we list the development of predictive in-home health monitoring systems as an important potential future research direction, especially considering the rapid technological advancements and their demonstrated benefits in proactive healthcare management.

\begin{tcolorbox}[colback=white!98!blue,colframe=blue!40!black]
\textbf{Recommendation 4:} Developing prediction/forecasting systems can lead to early detection and intervention, and thus better management of chronic conditions, making them a crucial area of research for improving patient outcomes and quality of care.
\end{tcolorbox}

\subsection{Personalized and adaptive systems}
The majority of existing in-home health monitoring systems provide standardized monitoring and intervention protocols that may not necessarily address individual needs and preferences. Therefore, it is important to develop personalized and adaptive in-home health monitoring systems by using data from various sources to provide tailored healthcare that is specific to each patient's unique needs \cite{amjad2023review}. Not only should personalized monitoring systems focus on tailoring healthcare to the unique needs of each individual but also on adapting to changes in their health status over time and with changing ambient conditions \cite{sebri2020introduction}. This individualized approach is especially crucial for diseases that exhibit a wide range of symptoms and progress differently from patient to patient. Studies focusing on chronic conditions like diabetes and hypertension \cite{chatrati2022smart}, where patient conditions can vary considerably over time, are prime examples of where these systems can have a significant impact. More specifically, in cardiac monitoring systems \cite{wang2023ecg}, the ability to adapt to changes in HR, rhythm, and other cardiovascular markers is key to providing more effective and personalized interventions. As another example, in the area of cancer treatment monitoring, these systems can play a pivotal role by optimizing treatment schedules and dosages based on real-time patient data~\cite{fu2022short}. They can accommodate the variability in patient responses to various forms of therapy. Furthermore, the management of neurological disorders \cite{miao2018wearable} also benefits from the adaptability of these systems. They can adjust therapeutic interventions in response to changes in symptoms or the progression of the condition. 
Recent advancements in DL such as continual learning \cite{molahasani2023continual,molahasani2023can} and personalized federated learning \cite{tan2022towards,fallah2020personalized} provide promising tools to address the personalization needs of in-home health monitoring systems. Through investing in research and development in such areas, researchers can enrich the quality of care received by subjects through personalized and adaptive systems.

\begin{tcolorbox}[colback=white!98!blue,colframe=blue!40!black]
\textbf{Recommendation 5:} The development of personalized and adaptive systems is necessary to further enhance the capabilities of in-home health monitoring systems' and ensure that patients receive the most tailored and up-to-date diagnoses and interventions possible.
\end{tcolorbox}

\subsection{More in-the-wild data and real-world deployment}
Many current datasets used in home-health monitoring systems are collected in controlled laboratory environments, which are limited in size and diversity of subjects, sensor variations, collection protocols, and ambient conditions~\cite{lara2012survey,meng2020recent}. Furthermore, patients may find existing sensors bulky and inconvenient, leading to non-compliance with monitoring protocols. These issues, among others, lead to many health-related systems being trained, evaluated, and consequently adept to controlled and relatively clean data, which can limit their effectiveness when applied to real-world, in-the-wild scenarios.
It is therefore crucial to collect large datasets in in-the-wild settings with which to train DL methods for health monitoring. To this end, developing effective mechanisms for data collection and labeling by a large number of participants in real-world settings is an important endeavor. Moreover, from a DL perspective, focusing on noisy \cite{djenouri2024spatio} and imbalanced \cite{wu2020fedhome} representation learning approaches can lead to significant improvements in robustness and deployability.

\begin{tcolorbox}[colback=white!98!blue,colframe=blue!40!black]
\textbf{Recommendation 6:} Curation and collection of large in-the-wild datasets, and prioritizing the usage of such data for developing in-home health monitoring models is a critical line of research that requires further attention. Furthermore, focusing on DL methods with the ability to process noisy, highly varied, and imbalanced data should be a priority in order to advance the possible deployment of accurate and reliable in-home health monitoring systems.
\end{tcolorbox}

\subsection{Large language models}  

Recently, large language models (LLMs)~\cite{hoffmann2022training}, such as BERT~\cite{devlin2018bert}, GPT~\cite{radford2019language}, and PaLM~\cite{chowdhery2022palm}, have shown remarkable performances on various tasks of natural language and multimodal understanding. These models are at the core of recent chatbots that can seamlessly communicate with individuals.  
Given the recent advances in conversational AI systems, we believe the integration of LLMs into in-home health monitoring systems can serve several critical purposes. A key application is the development of voice-based interaction models using LLMs \cite{yang2023talk2care}. These models can be adapted to recognize and respond to individual voices, offering personalized services such as medication reminders and health tracking, which is particularly beneficial for patients. This aligns with the need for individualized/personalized patient care highlighted in our survey. Furthermore, LLMs play a significant role in enhancing communication within telehealth services. Their ability to transcribe and analyze remote consultations makes them highly effective in improving the quality of remote healthcare delivery 
\cite{fang2020longitudinal}. In addition to these applications, we also believe such systems have the potential to contribute to in-home mental health systems. Through conversational inputs, LLMs can aid in identifying early signs of mental health disease \cite{malhotra2024xai,huq2022dialogue} and support patients with dementia and other cognitive diseases \cite{ruggiano2021chatbots}.

\begin{tcolorbox}[colback=white!98!blue,colframe=blue!40!black]
\textbf{Recommendation 7:} LLMs have enormous potential in the context of in-home health monitoring systems. These systems can be used to provide health-related cues, converse with users, and more, which can improve patient outcomes and enhance the quality of care. Therefore, we list this field as an important emerging research direction.
\end{tcolorbox}

\subsection{Less explored application areas}
Some application areas of in-home health monitoring systems have been more thoroughly explored, such as monitoring the elderly population \cite{kim2022home,moyle2021effectiveness,fares2021directing,wang2018leveraging}. However, other areas have received less attention in the literature. For example, monitoring infants and young children poses unique challenges that require further investigation.
As stated in \cite{hussain2019intelligent}, the integration of these systems into mobile platforms, particularly smartphones, presents a significant improvement in the quality of life in infants. Moreover, there is a need to study the long-term effects of such systems on parents and their behavioral changes toward providing effective health-related interventions for their babies \cite{bonafide2018accuracy}. One potential effect is the development of stress and anxiety in the case of false alarms, which may result in unnecessary medical interventions. As a result, this field is an interesting direction that must be explored. 
As another example of less explored areas, existing monitoring systems have not fully addressed some specific diseases and conditions, and research efforts should focus on developing systems to meet the needs of these patients \cite{pantelopoulos2009survey,titler2008evidence}. This will facilitate the development of more equitable and comprehensive health monitoring systems, ensuring that individuals with rare health conditions are not excluded. For example, rare genetic and neurological disorders \cite{chirra2019telemedicine} that affect a small subset of the population require tracking disease progression and providing personalized care. Similarly, diseases with non-specific symptoms or those that are difficult to diagnose, such as chronic fatigue syndrome or Lyme disease, can be diagnosed and treated by continuously monitoring and tracking the individual's health data.

\begin{tcolorbox}[colback=white!98!blue,colframe=blue!40!black]
\textbf{Recommendation 8:} It is crucial to continue to carry out research and development in areas that have received less attention, such as monitoring infants and young children, as well as specific diseases and conditions that have not yet been fully addressed by existing monitoring systems, to ensure that patients receive equitable and high-quality care.
\end{tcolorbox}

\section{Concluding Remarks}
\label{Sec: con}
In this paper, we provide a comprehensive review of current ubiquitous in-home health monitoring systems. We begin by identifying three main components of these systems which are sensing technologies, communication technologies, and intelligence and computing techniques. We then discuss each component in detail and highlight their recent advancements. Moreover, we review the applications of ubiquitous in-home health monitoring systems and explore how these systems can be utilized to monitor individuals' health. Our findings indicate that although there has been considerable progress in developing in-home health monitoring systems, there are still several limitations that need to be addressed. Finally, we point out some of these challenges and make 8 recommendations for future research directions. We believe this survey will be a valuable resource for researchers active in the field of ubiquitous in-home health monitoring.

 \bibliographystyle{ACM-Reference-Format}
\bibliography{sample-base}





\end{document}